\documentclass[usenatbib]{mn2e}
\usepackage{graphicx,times}
\usepackage{bm}
\usepackage{epsf}
\usepackage{wick}

\newcommand{\beq}{\begin{equation}}
\newcommand{\eeq}{\end{equation}}
\newcommand{\be}{\begin{equation}}
\newcommand{\ee}{\end{equation}}

\newcommand{\beqa}{\begin{eqnarray}}
\newcommand{\eeqa}{\end{eqnarray}}
\newcommand{\n}{\noindent}
\newcommand{\oh}{\hat \Omega}
\newcommand{\calC}{{\cal C}}
\newcommand{\nn}{{\nonumber}}
\newcommand{\calX}{{\cal X}}
\newcommand{\calY}{{\cal Y}}
\newcommand{\calZ}{{\cal Z}}
\newcommand{\calU}{{\cal U}}
\newcommand{\calV}{{\cal V}}
\newcommand{\calW}{{\cal W}}

\newcommand{\re}{\mathrm {Real}}

\def\qtwo{\qquad\qquad}

\def\ben{\begin{eqnarray}}
\def\een{\end{eqnarray}}

\def\myK{{\cal K}}

\def\bk{{\bf k}}
\def\bK{{\bf K}}

\def\2p{{(2\pi)^2}}

\def\la{{\langle}}

\def\tri{{T^{\calU_{l_a}\calV_{l_b}}_{\calW_{l_c}\calX_{l_d}}}}
\def\trip{{T^{\calU'_{l'_a}\calV'_{l'_b}}_{\calW'_{l'_c}\calX'_{l'_d}}}}
\def\dtri{{T^{\calU_{l}\calV_{l_b}}_{\calW_{l_c}\calX_{l_d}}}}

\date{\today,~ $ $Revision: 0.9 $ $}

\begin{document}

\onecolumn

\title[Primordial Non-Gaussianity and CMB Observations]
{Primordial Non-Gaussianity from a Joint Analysis of Cosmic
Microwave Background Temperature and Polarization}

\author[D. Munshi et al.]
{Dipak Munshi$^{1,2}$, Peter Coles$^{1}$, Asantha Cooray$^{3}$, Alan Heavens$^{2}$, Joseph Smidt$^{3}$\\
$^{1}$School of Physics and Astronomy, Cardiff University, Queen's Buildings, 5, The Parade, Cardiff, CF24 3AA, UK \\
$^{2}$Scottish Universities Physics Alliance (SUPA),~ Institute for Astronomy, University of Edinburgh,
Blackford Hill,  Edinburgh EH9 3HJ, UK\\
$^{3}$ Department of Physics and Astronomy, University of California, Irvine, CA 92697, USA}

\maketitle

\begin{abstract}
We explore a systematic approach to the analysis of primordial
non-Gaussianity using fluctuations in temperature and polarization
of the Cosmic Microwave Background (CMB). Following  Munshi \&
Heavens (2009), we define a set of power-spectra  as compressed
forms of the bispectrum and trispectrum derived from CMB temperature
and polarization maps; these spectra compress the information
content of the corresponding full multispectra and can be useful in
constraining early Universe theories. We generalize the standard
pseudo-$C_l$  estimators in such a way that they apply to these
spectra involving both spin-$0$ and spin-$2$ fields, developing
explicit expressions which can be used in the practical
implementation of these estimators. While these estimators are
suboptimal, they are nevertheless unbiased and robust hence can
provide useful diagnostic tests at a relatively small
computational cost. We next consider approximate inverse-covariance
weighting of the data and construct a set of near-optimal estimators
based on that approach. Instead of combining all available
information from the entire set of mixed bi- or trispectra, i.e
multispectra describing both temperature and polarization
information, we provide analytical constructions for individual
estimators, associated with particular multispectra. The bias and
scatter of these estimators can be computed using Monte-Carlo
techniques. Finally, we provide estimators which are completely
optimal for arbitrary scan strategies and involve inverse covariance
weighting; we present the results of an error analysis performed
using a Fisher-matrix formalism at both the one-point and two-point
level.
\end{abstract}

\begin{keywords}: Cosmology-- Cosmic microwave background-- large-scale structure
of Universe -- Methods: analytical, statistical, numerical
\end{keywords}

\section{Introduction}

The Cosmic Microwave Background (CMB) has the potential to provide
cosmologists with the cleanest statistical characterization of
primordial fluctuations. In most early universe studies the
primordial fluctuations are assumed to be nearly Gaussian, but the
quest for an experimental detection of primordial non-Gaussian is of
considerable importance.  Early observational work on the bispectrum
from COBE \citep{Komatsu02} and MAXIMA \citep{Santos} was followed
by much more accurate analysis using data from the Wilkinson
Microwave Anisotropy Probe
(WMAP)\footnote{http://map.gsfc.nasa.gov/}
\citep{Komatsu03,Crem07a,Spergel07}.With the recent claim of a
detection of non-Gaussianity \citep{YaWa08} in the 5-year WMAP data,
interest in non-Gaussianity has received a tremendous boost.
However, these, and most other, studies of primordial
non-Gaussianity focus primarily on data relating to temperature
anisotropies whereas inclusion of E-polarization data could, in
principle, increase the sensitivity of such tests. Ongoing surveys,
such as that derived from the Planck Surveyor\footnote
{http://www.rssd.esa.int/index.php?project=Planck}, will throw more
light on this question both by improving sensitivity to both
temperature and polarization fluctuations. It is the primary purpose
of this paper to extend a number of previously obtained analytical
results in order to design a new set of diagnostic tools for the
analysis of primordial non-Gaussianity using using both temperature
and polarization data.

The CMB polarization field can be decomposed into a gradient part
with even parity (the ``electric" or $E$-mode) and a curl part with
odd parity (the ``magnetic" or $B$-mode), with important
ramifications for its interpretation in a cosmological setting.
Scalar (density) perturbations, at least in the linear regime, are
unable to generate $B$-mode polarization directly. Secondary effects
such as gravitational lensing can generate magnetic polarization
from an initial purely electric type, but only on relatively small
angular scales. Tensor (gravitational-wave) perturbations, however,
can generate both $E$-mode and $B$-mode on large scales. Detection
of $B$-mode polarization on large angular scales is therefore, at
least in principle, a tell-tale signature of the existence of
gravitational waves predicted to be generated during inflation. The
amplitude of these tensor perturbations also contains information
relating to the energy scale of inflation and thus has considerable
power to differentiate among alternative models of the early
Universe.

A number of experiments therefore are either being planned or
currently underway to characterize the polarized CMB sky, including
Planck. The primary statistical tools used currently to characterize
the stochastic polarized component are the power spectra of the $E$
and $B$-mode contributions; these have been studied in the
literature in great detail for various survey strategies,
non-uniform noise distribution, and partial sky coverage \citep{Hiv02}.
Most studies in this vein involve a (computationally extensive)
maximum likelihood analysis or a quadratic estimator, which have
limited ability to handle maps with large numbers of pixels, or the
so-called pseudo-$C_l$s estimator (PCL) which uses a heuristic
weighting of various pixels, instead of an optimal or near-optimal
one.

Information about non-Gaussianity, however, is not contained in the
power-spectrum $C_l$, however it is estimated. A recent study by
\citet{Munshi_kurt} extended the PCL approach to the study of
non-Gaussianity by taking into account higher-order spectra, as well
as clarifying the relationship of the estimators obtained to the
optimal ones. This was done for temperature only, so one of the aims
of this paper is to generalise this early work to take into account
both temperature and polarization.

Approaches based on PCL do not require detailed theoretical
modelling of the target bispectrum. Optimal estimators on the other
hand, work using a matched-filter approach which does  require
analytical modelling. The simplest inflationary models - based on a
single slowly-rolling scalar field - are associated with a very
small level  of non-Gaussianity
\citep{Salopek90,Salopek91,Falk93,Gangui94,Acq03,Mal03}; see
\citet{Bartolo06} and references there in for more details. More
elaborate variations on the inflationary theme, such as those
involving multiple scalar fields \citep{Lyth03}, features in the
inflationary potential, non-adiabatic fluctuations, non-standard
kinetic terms, warm inflation \citep{GuBeHea02,Moss}, or deviations
from Bunch-Davies vacuum can all lead to much higher level of
non-Gaussianity.

Generally speaking, the apparatus used to model primordial
non-Gaussianity has focused on a phenomenological `{\em local}
$f_{NL}$' parametrization in terms of the perturbative non-linear
coupling in the primordial curvature perturbation \citep{KomSpe01}:
\begin{equation}
\Phi(x) = \Phi_L(x) + f_{NL} [\Phi^2_L(x) - \langle \Phi^2_L(x) \rangle ] +
g_{NL}\Phi^3_L(x)
+ \dots ,
\end{equation}
where $\Phi_L(x)$ denotes the linear Gaussian part of the Bardeen
curvature \citep{Bartolo06} and $f_{NL}$ is the non-linear coupling parameter.
A number of models have non-Gaussianity which can be approximated by
this form. The leading order non-Gaussianity therefore is at the
level of the bispectrum or, equivalent, at the three-point level in
configuration space. Many studies involving primordial
non-Gaussianity have used the bispectrum, motivated by the fact that
it contains all the information about $f_{NL}$ \citep{Babich}. This
has been extensively studied
\citep{KSW,Crem03,Crem06,MedeirosContaldi06,Cabella06,Liguori07,SmSeZa09},
with most of these measurements providing convolved estimates of the
bispectrum. Optimized 3-point estimators were introduced by
\citet{Heav98}, and have been successively developed
\citep{KSW,Crem06,Crem07b,SmZaDo00,SmZa06} to the point where an
estimator for $f_{NL}$ which saturates the Cramer-Rao bound exists
for partial sky coverage and inhomogeneous noise \citep{SmSeZa09}.
Approximate forms also exist for {\em equilateral} non-Gaussianity,
which may arise in models with non-minimal Lagrangian with
higher-derivative terms \citep{Chen06,Chen07}. In these models, the
largest signal comes from spherical harmonic modes with
$\ell_1\simeq \ell_2 \simeq \ell_3$, whereas for the local model,
the signal is highest when one $\ell$ is much smaller than the other
two. Moreover, the covariances associated with these power-spectra
can be computed analytically for various models, thereby furnishing
methods to test simulation pipeline in a relatively cost-effective
way.

As we mentioned above, most analysis of the CMB bispectrum take into
account only temperature information because of lower sensitivity
associated with polarisation measurements. However, the overall
signal-to-noise ratio of a detection of non-Gaussianity can in
principle be improved by incorporating $E$-type polarisation
information in a joint analysis. Several ground-based or future experiments
\footnote{See e.g. http://cmbpol.uchicago.edu/workshops/path2009/abstracts.html for various
ongoing and planned surveys.} have either started or will be measuring $E$-polarisation, and ongoing
all-sky experiments such as  Planck will certainly improve the
signal to noise. Moreover, there are many planned experiments which
will survey the sky with even better sensitivity; see e.g.
\citep{Bau09} for the CMBPol mission concept study. 
It is therefore
clearly timely to update and upgrade the available estimators which
can analyze non-Gaussianity including both temperature and
polarization. The main question is to how to do it optimally.

The fast estimator introduced initially by \cite{KSW} can handle
temperature and polarization data. Extension of this estimator to
take into account a linear term, which can reduce the variance in
the presence of partial sky coverage, was introduced later in
\cite{Crem06}. In the absence of a linear term the scatter
associated with the estimator increases at high -$\ell$. A more
general approach, that includes inverse variance weighting of the
data, was introduced by \cite{SmZa06} and can make the estimator
optimal. However most of these estimators tend to compress all
available information into a single point. This has the advantage of
reducing the scatter in the estimator but it throws away a great
deal of potentially relevant information.

Recently \citet{MuHe09} have extended this approach by devising a
method which can map the bispectrum,  or even higher-order spectra
such as the trispectrum, to associated power-spectra; these power
spectra are related to the {\it cumulant correlators} used in the
context of projected galaxy surveys \citep{MuMeCo99,SS99}. These compressed
spectra do not contain all the information contained in the
multispectra but they certainly do carry more information than a
single number, and for certain purposes, such as $f_{NL}$ estimation, can be completely optimal.

In this paper we extend the analysis of
\citet{MuHe09}  to the case of a joint analysis of temperature and
$E$-mode polarization data. We provide realistic estimators for
power spectra related to {\em mixed} bispectra and trispectra (i.e.
multispectra incorporating and describing both temperature and
polarization properties of the radiation field) in the presence of
partial sky coverage. We then generalize these results to the case
of optimized estimators which can handle all realistic complications
by an appropriate optimal weighting of the data. We also provide a
detailed analysis of pseudo-C$_{\ell}$ based estimators. Though
suboptimal, these estimators are unbiased and turn out to be
valuable diagnostics for analyzing large number of simulations very
quickly.

The plan of the paper is as follows. In \textsection2 we describe
the basic results needed for the PCL analysis. We then introduce the
power spectrum associated with the mixed bispectrum. The formalism
is sufficiently general to handle both $E$ and $B$-mode, but for
simplicity and practical relevance we give results for the $E$-mode
only. These results are next generalized in \textsection4 to
near-optimal estimators which were introduced by \citet{KSW}. The
optimization depends on matched-filtering techniques based on
theoretical modeling of primordial non-Gaussianity presented in
\textsection3 and relies on Monte-Carlo standardization of bias and
scatter. Next, we consider the approach where various fields are
weighted by inverse covariance matrices to make them optimal in
\textsection5. Though this approach does not rely on Monte-Carlo
simulations, accurate modelling of inverse covariance matrix can be
computationally expensive and can only be achieved for maps with
relatively low resolution.

\section{Method of Direct Inversion or Pseudo $C_{\ell}$  Approach}

Maximum likelihood analysis techniques, or related quadratic
estimators, are generally used to estimate the power spectrum from a
given cosmological data set such as a map of the CMB. This approach
relies on (optimal) inverse-covariance weighting of the data which,
unfortunately, is virtually impossible to implement in practice for
large data sets. However, faster {\em direct} inversion techniques
with heuristic weighting schemes are faster and are typically
employed for estimation of power spectrum from cosmological data
\citep{Hiv02}. The alternative,  PCL, technique discussed above can,
however, be used to study the power spectrum related to the bispectrum.
This method is not optimal, but remains unbiased and can act as a
precursor to more computationally expensive studies using various
optimal estimators.

This section is devoted to the generalization of the pseudo-$C_l$
(PCL) results to the {\em skew-spectrum} associated with spinorial
fields such as  polarization radiation. The results we  shall derive
are valid for all-sky coverage as well as for partial sky coverage;
we relate the two using a coupling matrix which depends on the power
spectrum associated the mask describing the coverage. We provide
results for the coupling matrix for various generic combinations of
spin-$0$ and spin-$2$ fields. These results extend the results
obtained previously for the temperature-only case (spin-$0$).

\subsection{The Power Spectrum associated with the Bispectrum}

We start by defining the complex field  $P_{\pm}(\oh)$ constructed
from the Stoke's parameters $Q(\oh)$ and $U(\oh)$. \beq P_{\pm}(\oh)
= Q(\oh) \mp iU(\oh) \eeq The fields denoted $P_{\pm}(\oh)$ are
spin-$2$ (tensor) fields, whereas the temperature field can be
represented as a spin-$0$ (scalar) field on the surface of the sky.
The appropriate multipole expansion of these objects is performed
using the {\it spin-$2$ spherical harmonics}, ${}_{\pm 2}Y_{lm}$, as
basis functions. We will denote the spin harmonics of spin $s$ with
${}_sY_{lm}(\oh)$ on the surface of a unit sphere.

\beq
P_{\pm}(\oh) = \sum_{lm} {}_{\mp2}Y_{lm}(\oh) (E_{lm} \pm B_{lm});  ~~~~~ [P_{\pm}]_{lm} = (E_{lm} \pm B_{lm}).
\eeq

\n The terms $E_{lm}$ and $B_{lm}$  are the harmonic components of
{\it Electric} and {\it Magnetic} components respectively. It should
be clear that the fields $P_{\pm}$, constructed from $Q$ and $U$,
correspond to spin $\mp 2$ respectively. 

Estimation of power spectrum
for the $E$ and $B$ fields from experiments with partial sky
coverage is of great importance for cosmological experiments and has
attracted a great deal of interest.

We start by considering two fields on the surface of a sphere ${\cal
X}(\oh)$ and ${\cal Y}(\oh)$, which are respectively of spin $x$ and
$y$. These objects can be $P_{\pm}$ or $\delta T$. Resulting product
fields such as  $P_{+}P_{-}$ can be of spin zero, while those like
$P_{+}^2$ and $P_{-}^2$ are of spin $-4$ and $+4$ respectively. The
results can also involve fields such as $\delta_T P_{+}$ which is a
spin $-2$ field. All these spin-$s$ fields can be decomposed using
spin-harmonics ${}_sY_{lm}(\oh)$ given above. These spin harmonics
are generalization of ordinary spherical harmonics $Y_{lm}(\oh)$
which are used to decompose the spin-$0$ (or scalar) functions e.g.
$\delta T$ defined over a surface of the celestial sphere.
Throughout we will be using lower case symbols $x,y$ to denote the
spins associated with the corresponding fields denoted in italics.
The fields that are constructed from various powers of $P_{+}$ and
$P_{-}$ can be expanded in terms of the corresponding spin-harmonics
basis ${}_sY_{lm}(\oh)$. The product field such as $[{\cal
X}(\oh){\cal Y}(\oh)]$ which is of spin $x+y$ can therefore be
expanded in terms of the harmonics ${}_{x+y}Y_{lm}(\oh)$. The
following relationship therefore expresses the harmonics of the
product field in terms of the harmonics of individual fields.

\beqa
[{\cal X}(\oh){\cal Y}(\oh)]_{lm} &=& \int d\oh {\cal X}(\oh){\cal Y}(\oh)[{}_{x+y}Y_{lm}^*(\oh)] \\
&&  \sum_{l_im_i} \calX_{l_1m_1} \calY_{l_2m_2} \int [{}_xY_{l_1m_1}(\oh)][{}_yY_{l_2m_2}(\oh)]
[{}_{x+y}Y_{lm}^*(\oh)] d\oh;  ~~~~ \calX, \calY \in P_{\pm}  \\
&&  = \sum_{l_im_i}\calX_{l_1m_1} \calY_{l_2m_2} I_{l_1l_2l}\left ( \begin{array} { c c c }
     l_1 & l_2 & l \\
     x & y & -(x+y)
  \end{array} \right )\left ( \begin{array} { c c c }
     l_1 & l_2 & l \\
     m_1 & m_2 & -m
  \end{array} \right ); ~~~~ I_{l_1l_2l} = \sqrt {(2l_1+1)(2l_2+1)(2l+1) \over 4\pi }.
\label{eq:decompose}
\eeqa

\n The expression derived above is valid for all-sky coverage; it
will be generalized later to take into account arbitrary partial sky
coverage. We have assumed that any noise contamination is Gaussian,
so that it will not contribute to this non-Gaussianity statistic. To
define the associated power spectra we can write:

\beq
C_l^{\calX \calY, \calZ} = {1 \over 2l+1} \sum_m [XY]_{lm} Z^*_{lm} =
\sum_{l_1l_2} B^{\calX\calY\calZ}_{l_1l_2l} \sqrt {   {(2l_1+1)(2l_2+1) \over ( 2l+1 )}  }
\left ( \begin{array} { c c c }
     l_1 & l_2 & l \\
     x & y & -(x+y)
  \end{array} \right ).
\label{eq:bispec}
\eeq

\n
Here we have introduced the bispectrum  $B^{\calX\calY\calZ}_{l_1l_2l}$ which
can be related to the harmonics of the relevant fields by the following equation:

\beq
B_{l_1l_2l_3}^{\calX\calY\calZ} = \sum_{m_1m_2m_3} \langle\calX_{l_1m_1} \calY_{l_2m_2} \calZ_{l_3m_3} \rangle_c
\left ( \begin{array} { c c c }
     l_1 & l_2 & l_3 \\
     m_1 & m_2 & m_3
  \end{array} \right ).
\eeq

The matrices denote the Wigner-$3j$ symbols \citep{Ed68} which are
only non-zero when the quantum numbers $l_i$ and $m_i$ satisfy
certain conditions. For $x=y=0$ these results generalize those obtained in
\citet{Cooray01} valid for the case of temperature; see also
\cite{ChenSzapudi} for related discussions. The result derived above
is valid for a general $E$ and $B$ type polarization field. The
contribution from $B$-type magnetic polarization is believed to be
considerably smaller than the $E$-type electric polarization. The
results derived above will simplify considerably if we ignore the
$B$ type polarisation field in our analysis. In general the
power-spectra described above will be complex functions, though the
real and imaginary parts can be separated by considering different
components of the bispectrum. However, if we assume that the
magnetic part of the polarization is zero the following equalities
will hold:

\beq
B_{l_1l_2l_3}^{\delta T \delta T \delta T}  =   B_{l_1l_2l_3}^{TTT};  \\
B_{l_1l_2l_3}^{P_{\pm}\delta T \delta T}  =  B_{l_1l_2l_3}^{ETT};  \\
B_{l_1l_2l_3}^{P_{\pm}P_{\pm}\delta T}  = B_{l_1l_2l_3}^{EET}; \\
B_{l_1l_2l_3}^{P_{\pm}P_{\pm}P_{\pm}}  =  B_{l_1l_2l_3}^{EEE}.  \\
\eeq

The result detailed above is valid for all-sky coverage. Clearly,
for our results to be relevant in practical applications, we need to
add a Galactic mask. If we consider the masked harmonics associated
where the (arbitrary) mask $w(\oh)$ then, expanding the product
field in the presence of the mask, we can write:

\beqa
&& [{\cal X}(\oh){\cal Y}(\oh)w(\oh)]_{lm} =
\sum_{l_im_i;l_am_a}  \calX_{l_1m_1} \calY_{l_2m_2} w_{l_am_a} \int [{}_xY_{l_1m_1}(\oh)][{}_yY_{l_2m_2}(\oh)] [Y_{l_am_a}]
[{}_{x+y}Y_{lm}^*] d\oh \nn \\
&& = \sum_{l_im_i}\sum_{l_am_a} (-1)^{l'}\calX_{l_1m_1} \calY_{l_2m_2} w_{l_am_a} I_{l_1l_2l} \left (  \begin{array} { c c c }
     l_1 & l_2 & l' \\
     x & y & -(x+y)
  \end{array} \right ) \left (\begin{array} { c c c }
     l_1 & l_2 & l' \\
     m_1 & m_2 & -m'
  \end{array} \right ) \int [{}_{-(x+y)}Y_{l'm'}^*] Y_{l_am_a}[{}_{x+y}Y_{lm}^*] \nn \\
&& \qquad \qquad \qquad\qquad = \sum_{l_im_i} (-1)^{l+l'}\sum_{l_am_a} \calX_{l_1m_1} \calY_{l_2m_2} w_{l_am_a} I_{l_1l_2l'}I_{l'l_al}
  \left (  \begin{array} { c c c }
     l_1 & l_2 & l' \\
     x & y & -(x+y)
  \end{array} \right ) \left (\begin{array} { c c c }
     l_1 & l_2 & l' \\
     m_1 & m_2 & -m'
  \end{array} \right ) \nn \\
&& \qquad \qquad \qquad\qquad \qquad \qquad\qquad \qquad \qquad\qquad \qquad \qquad
 \times \left (  \begin{array} { c c c }
     l' & l_a & l \\
     (x+y) & 0 &  -(x+y)
  \end{array} \right ) \left (\begin{array} { c c c }
     l' & l_a & l \\
     m' & m_a & -m
  \end{array} \right ).
\eeqa

In simplifying the relations derived in this section we have used
the relationship Eq.(\ref{eq:harmonics1}) and
Eq.(\ref{eq:harmonics2}). We have also used the fact that
${}_sY_{lm}^* = (-1)^{m+s}Y_{l,-m}$, where ${}^*$ denotes the
complex conjugate. Similarly, we can express the pseudo-harmonics of
the field $\calZ(\oh)$ observed with the same mask in terms of its
all-sky harmonics.

\beqa
[{\cal Z}(\oh)w(\oh)]_{lm} &=& \int d\oh [{\cal Z}(\oh)w(\oh)] [{}_{z}Y_{lm}^*]
=\sum_{l_im_i} \calZ_{l_3m_3} w_{l_bm_b}  \int [{}_zY_{l_3m_3}(\oh)] [Y_{l_bm_b}(\oh)]
[{}_{z}Y_{lm}^*] d\oh \nn \\
&& = \sum_{l_3m_3}\sum_{l_bm_b} \calZ_{l_3m_3} w_{l_bm_b} I_{l_3l_bl} \left (  \begin{array} { c c c }
     l_3 & l_b & l \\
     z & y & -z
  \end{array} \right ) \left (\begin{array} { c c c }
     l_3 & l_b & l \\
     m_3 & m_b & -m
  \end{array} \right ).
\eeqa

The harmonics of the composite field $[{\cal X}(\oh){\cal Y}(\oh)]$
when constructed on a partial sky are also  functions of the
harmonics of the mask used, $w_{lm}$. The simplest example of a mask
would be $w=1$ within the observed part of the sky and $w=0$
outside. For a more complicated mask, the harmonics  $w_{lm}$ are
constructed out of spherical harmonics transforms. We also need to
apply the mask to the third field which we will be using in our
construction of the power spectrum related to the bispectrum (which
we sometimes refer to as the skew-spectrum) associated with these
three fields. We will take the masks in each case to be the same,
but the results could be very easily generalized for two different
masks. The {\it pseudo} power spectrum is constructed from the
masked harmonics of the relevant fields in the form of a
cross-correlation power spectrum:

\beqa
\tilde C_l^{\calX\calY,\calZ} = {1 \over 2l+ 1}
\sum_{m=-l}^{l} [{\cal X}(\oh){\cal Y}(\oh)w(\oh)]_{lm} [{\cal Z}(\oh)w(\oh)]^*_{lm}.
\eeqa

\n The pseudo power spectrum $\tilde C_l$ is thus a linear
combination of the modes of its all-sky counterpart $C_l$. In this
sense, masking introduces a coupling of modes of various order which
is absent in the case of all-sky coverage. The matrix $M_{ll'}$ encodes
the information regarding the mode-mode coupling and depends on the 
power spectrum of the mask $w_l$. For surveys with partial sky
coverage the matrix is not invertible which signifies the loss of
information due to masking. Therefore, binning of the pseudo
skew-spectrum or any higher-order version based on, e.g., kurtosis,
may be necessary before it can be inverted, which leads to the recovery
of (unbiased) binned spectrum.

\beq
\tilde C_l^{\calX\calY,\calZ} = \sum_{l'} M^{xy,z}_{ll'} C_{l'}^{\calX\calY,\calZ}.
\eeq

\n The all-sky power spectrum $C_l^{\calX\calY,\calZ}$ can be
recovered by inverting the equation related to the accompanying
bispectrum by Eq.(\ref{eq:bispec}) discussed above. The
mode-mode coupling matrix depends not only on the power spectrum
$|w_{l}|$ of the mask but also on the spin associated with the
various fields which are being probed in construction of
skew-spectrum:

\beq
M_{ll'}^{xy,z} = {1 \over 4\pi }\sum_{l_a} (2l'+1)(2l_a+1) \left ( \begin{array}{ c c c }
     l & l_a & l' \\
     x+y & 0 & -{(x+y)}
  \end{array} \right )
\left ( \begin{array} { c c c }
     l & l_a & l' \\
     z & 0 & -z
  \end{array} \right ) |w_{l_a}|^2
\eeq

This expression reduces to that of the temperature bispectrum if we set
all the spins to be zero $x=y=z=0$, in which case the coupling of
various modes due to partial sky coverage only depends on the power
spectrum of the mask. In case of temperature-only (spin-$0$)
analysis we have seen that the mode-mode coupling matrix do not
depend on the order of the statistics. The {\it skew-spectrum} or
the power spectrum related to bispectrum, as well as its higher
order counterparts, such as the power spectrum related to
trispectrum, can all have the same mode-mode coupling matrix in the
presence of partial sky coverage. However, this is not the case for
power-spectra related to the polarization multispectra; that depends
on the spin of various fields used to construct the bispectrum. It
is customary to define a single number associated with each of these
bispectra. The {\it skewness} is a weighted sum of the power
spectrum related to the bispectrum $S_3^{\calX\calY,\calZ} = \sum_l
(2l+1) C_l^{\calX\calY,\calZ}$.

We list the specific cases of interest below. The relations between
the bispectra and the associated power spectra generalize cases
previously obtained where only the temperature bispectrum was
considered \citep{Cooray01}.

\beqa
&& C_l^{TT,E} =
\sum_{l_1l_2} B^{TTE}_{l_1l_2l} \sqrt {   {(2l_1+1)(2l_2+1) \over 4\pi ( 2l+1 )}  }
\left ( \begin{array} { c c c }
     l_1 & l_2 & l \\
     0 & 0 & 0
  \end{array} \right ) \\
&& M_{ll'}^{00,2} = {1 \over 4\pi }\sum_{l_a} (2l'+1)(2l_a+1) \left ( \begin{array}{ c c c }
     l & l_a & l' \\
     0 & 0 &  0
  \end{array} \right )
\left ( \begin{array} { c c c }
     l & l_a & l' \\
     2 & 0 & -2
  \end{array} \right ) |w_{l_a}|^2.
\label{eq:TTE}
\eeqa

\n
Next we can express the power spectrum $C_l^{TE,E}$ probing the mixed bispectrum $B^{TEE}_{l_1l_2l}$
by the following relation:

\beqa
&& C_l^{TE,E} =
\sum_{l_1l_2} B^{TEE}_{l_1l_2l} \sqrt {   {(2l_1+1)(2l_2+1) \over 4\pi ( 2l+1 )}  }
\left ( \begin{array} { c c c }
     l_1 & l_2 & l \\
     2 & 0 & -2
  \end{array} \right ) \\
&& M_{ll'}^{02,2} = {1 \over 4\pi }\sum_{l_a} (2l'+1)(2l_a+1) \left ( \begin{array}{ c c c }
     l & l_a & l' \\
     2 & 0 &  -2

  \end{array} \right )
\left ( \begin{array} { c c c }
     l & l_a & l' \\
     2 & 0 & -2
  \end{array} \right ) |w_{l_a}|^2.
\label{eq:TEE}
\eeqa

\n Similarly, for the case  $C_l^{EE,E}$, which probes the
bispectrum $B^{EEE}_{l_1l_2l}$, we  have the following expressions.

\beqa
&& C_l^{EE,E} =
\sum_{l_1l_2} B^{EEE}_{l_1l_2l} \sqrt {   {(2l_1+1)(2l_2+1) \over 4\pi ( 2l+1 )}  }
\left ( \begin{array} { c c c }
     l_1 & l_2 & l \\
     4 & 0 & -4
  \end{array} \right ) \\
&& M_{ll'}^{22,2} = {1 \over 4\pi }\sum_{l_a} (2l'+1)(2l_a+1) \left ( \begin{array}{ c c c }
     l & l_a & l' \\
     4 & 0 &  -4
  \end{array} \right )
\left ( \begin{array} { c c c }
     l & l_a & l' \\
     2 & 0 & -2
  \end{array} \right ) |w_{l_a}|^2.
\label{eq:EEE}
\eeqa

Other power spectra, such as $C_l^{EE,T}$ and those involving B-mode
polarization, can be expressed in terms of the underlying bispectrum
in a very similar manner.  It is likely that future high
sensitivity experiments can probe these power spectra individually.
As these are unbiased estimators they can also be useful in testing
simulation pipelines involving a large number of Monte-Carlo
realizations, which are routinely used for standardization and
characterization of data-analysis pipelines. The inversion of the
coupling matrix $M_{ll'}$ can require binning, in which case the
relevant binned coupling matrix $M_{bb'}$ in the case of small sky
coverage (such as ground-based and balloon-borne) surveys will lead
to recovery of binned power spectra $C_{l_b}$. The flat sky analogs
of these results can be useful in other cosmological studies,
including weak lensing observations, involving spin-2 fields
(mapping out the shear or $\gamma$ on the surface of the sky) that
covers a small fraction of the sky. We plan to present results of
such an analysis elsewhere (Munshi et al. 2010, in preparation).

\subsection{The Power spectrum associated with the Trispecrum}

The bispectrum represents the lowest-order deviation from
Gaussianity; the next highest order is the trispectrum. There are
many reasons for wanting to go beyond the lowest-order description.
Many studies relating to secondary anisotropies have shown that
ongoing surveys such as Planck can provide information about
non-Gaussianity beyond lowest order. These include the mode coupling
of CMB due to weak lensing of CMB as well as the other secondary
effects such as thermal Sunyaev-Zeldovich (tSZ) and kinetic
Sunyaev-Zeldovich effect (kSZ) \citep{CoorayHu}. With the recent
(claimed) detection of primordial non-Gaussianity in the CMB, there
has also been a renewed interest in detection of primordial
non-Gaussianity beyond lowest order with reasonable signal-to-noise.
Such studies will constrain the parameters $g_{NL}$ and $f_{NL}$
independently. The signal-to-noise for such constraints is  weaker
compared to the ones achieved using bispectrum, but with anticipated
future increases in sensitivity of ongoing as well as future planned
surveys such studies will play an important role in future. Future
21cm surveys  can also provide valuable information about the
trispectrum \citep{CooMe08}. A recent study by \citet{Munshi_kurt}
constructed an estimator for trispectrum which can work with
temperature data and which is optimized to detect primordial
non-Gaussianity.This statistic was also applied to WMAP 5-year data in \citep{Smidt2010}
to provide the first constraints on $\tau_{\rm NL}$ and $g_{\rm NL}$,
the third order corrections to primordial perturbations in
non-Gaussian models. We generalise and extend these works here to include both
temperature and polarizations data for independent or joint estimate
of such quantities.

\subsubsection{The Two-to-Two Power spectrum}

In constructing the power-spectrum related to the trispectrum, we
start with two fields $\cal U$ and $\cal V$ respectively on the
surface of the sphere. As before, we can take specific examples
where these fields are either $P_{\pm}$ or $\delta T$. We will keep
the analysis generic here, but will consider more specific examples
later on. The spins associated with various fields are denoted by
lower case symbols, i.e. $u$ and $v$. The product field now can be
expanded in terms of the spin harmonics of spin $u+v$, as was done
in Eq.(\ref{eq:decompose}). Similarly, decomposing the other set of
product field  we obtain $[\calW(\oh)\calX(\oh)]_{lm}$. We next
construct the the power spectrum associated with the trispectrum
from these harmonics:

\beq
C_l^{\calU\calV,\calW\calX} = {1 \over 2l +1} \sum_{m=-l}^l [\calU(\oh)\calV(\oh)]_{lm} [\calW(\oh)\calX(\oh)]_{lm}^*.
\eeq

\n This particular type of power spectrum associated with trispectra
has been studied extensively in the literature in the context of CMB
studies (see e.g. \cite{wl03}). It is one of the two degenerate power spectra
associated with trispectrum. After going through very similar
algebra outlined in the previous section we can express the
$C_l^{UV,WX}$ in terms of the relevant trispectra which it is
probing. The resulting expression, in the absence of any mask, takes
the following form:

\beqa
&& C_l^{\calU\calV,\calW\calX} = \sum_{l_1,l_2,l_3,l_4} T^{U_{l_1}V_{l_2}}_{W_{l_3}X_{l_4}}(l)
\left ( \begin{array}{ c c c }
     l_1 & l_2 & l \\
     u & v & -{(u+v)}
  \end{array} \right )
\left ( \begin{array} { c c c }
     l_3 & l_4 & l \\
     x & y & -{(x+y)}
  \end{array} \right )
\sqrt {   {(2l_1+1)(2l_2+1) \over ( 2l+1 )}  }
\sqrt {  (2l_3+1)(2l_4+1) \over (2l+1)  }; \\
&& \qquad \qquad \qquad \qquad \qquad \qquad \qquad \qquad  \calU,\calV,\calW,\calX \in P_{\pm}, \delta_T; \nn \\
&& \qquad \qquad \langle \calU_{l_1m_1}\calV_{l_2m_2}\calW_{l_3m_3}\calX_{l_4m_4}\rangle_c
= \sum_{LM}(-1)^M\left ( \begin{array}{ c c c }
     l_1 & l_2 & L \\
     m_1 & m_2 & -M
  \end{array} \right )
\left ( \begin{array}{ c c c }
     l_3 & l_4 & L \\
     m_3 & m_4 & M
  \end{array} \right)T^{U_{l_1}V_{l_2}}_{W_{l_3}X_{l_4}}(L)
\eeqa

\n In the presence of a completely general mask $w(\oh)$ the
pseudo-$C_{\ell}$s or PCLs will have to be modified. This involves
computing the spherical harmonics of the masked field
$\calU(\oh)\calV(\oh)w(\oh)$ and cross-correlating it against the
harmonics of $\calW(\oh)\calX(\oh)w(\oh)$.

\beq
\tilde C_l^{\calU\calV,\calW\calX} = {1 \over 2l +1} \sum_{m=-l}^l [\calU\calV~w]_{lm} [\calW\calX~w]_{lm}^*;
\qquad \qquad w_{l} = {1 \over 2l+1} \sum_{m=-l}^l w_{lm}w_{lm}^*.
\eeq

Again, the resulting PCLs are linear combinations of their all-sky
counterparts. The mixing matrix which encodes  information about the
mode mixing will depend on the power spectrum of the mask as well as
the spins of all four associated fields. The mixing matrix $M_{ll'}$
expressed in terms of the Wigner's $3j$ symbols takes the following
form:

\beq
M_{ll'}^{uv,xy} = {1 \over 4\pi }\sum_{l_a} (2l'+1)(2l_a+1) \left ( \begin{array}{ c c c }
     l & l_a & l' \\
     u+v & 0 & -{(u+v)}
  \end{array} \right )
\left ( \begin{array} { c c c }
     l & l_a & l' \\
     x+y & 0 & -(x+y)
  \end{array} \right ) |w_{l_a}|^2; \qquad \qquad u,v,x,y \in \pm2,0.
\eeq

\n
The pseudo-$C_\ell$s expressed as a linear combination of all-sky power spectra can now be expressed
using the following relationship:

\beq
\tilde C_l^{\cal UV,WX} = \sum_{l'} M_{ll'}^{uvw,x} C_{l'}^{\cal UV,WX}.
\eeq

For nearly complete sky surveys, and with proper binning, the mixing
matrix $M_{ll'}$ can be made invertible. This provides a unique way
to estimate all-sky  $C_{l'}^{UVW,X}$ and extract the information it
contains about the trispectra. For a given theoretical prediction,
the all-sky power spectra $C_{l'}^{UVW,X}$ can be analytically
computed. Knowing the detailed model of an experimental mask then
allows us to compute the observed $\tilde C_l^{UVW,X}$ accurately.
The results presented here generalizes those obtained in
\citep{Munshi_kurt} for the case of polarisation studies.

These results assumes generic field variables ${\cal X}(\oh)$,
${\cal Y}(\oh)$ which can have arbitrary spin associated to them. We
next specify certain specific cases where we identify three of the
fields ${\cal X, Y, Z} = \delta_T$ and ${\cal Z}= P_{+}= E$. The
other combinations can also be obtained in a similar manner.

\beqa
&& C_l^{TT,TE} = \sum_{l_1,l_2,l_3,l_4} T^{T_{l_1}T_{l_2}}_{T_{l_3}E_{l_4}}(l)
\left ( \begin{array}{ c c c }
     l_1 & l_2 & l \\
     0 & 0 & 0
  \end{array} \right )
\left ( \begin{array} { c c c }
     l_3 & l_4 & l \\
     2 & 0 & -2
  \end{array} \right )
\sqrt {   {(2l_1+1)(2l_2+1) \over ( 2l+1 )}  }
\sqrt {  (2l_3+1)(2l_4+1) \over (2l+1)  }; \\
&& M_{ll'}^{00,02} = {1 \over 4\pi }\sum_{l_a} (2l'+1)(2l_a+1) \left ( \begin{array}{ c c c }
     l & l_a & l' \\
     0 & 0 &  0
  \end{array} \right )
\left ( \begin{array} { c c c }
     l & l_a & l' \\
     2 & 0 & -2
  \end{array} \right ) |w_{l_a}|^2;
\eeqa

\n Other estimators for mixed trispectra involving different
combinations of $E$-polarization and temperature anisotropy
$\delta_T$ can be derived in a similar manner and can provide
independent information of corresponding trispectra. As before, we
have ignored the presence of $B$-type polarization in our analysis.
The presence of a non-zero $B$-mode can be dealt with very easily in
our framework but the resulting expressions will be more
complicated.

\begin{figure}
\begin{center}
{\epsfxsize=9. cm \epsfysize=8. cm {\epsfbox[22 184  590 605]{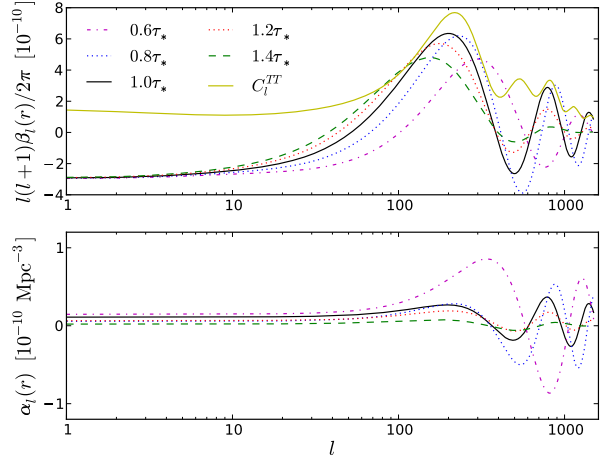}}}
\end{center}
\caption{$\alpha(r)$ (lower panel) and $\beta(r)$ (upper panel) for Temperature are plotted as a function of $l$. Plots are based on WMAP7 parameters \citep{WMAP7}}
\label{fig:abT}
\end{figure}

\begin{figure}
\begin{center}
{\epsfxsize=9. cm \epsfysize=8. cm {\epsfbox[22 184  590 605]{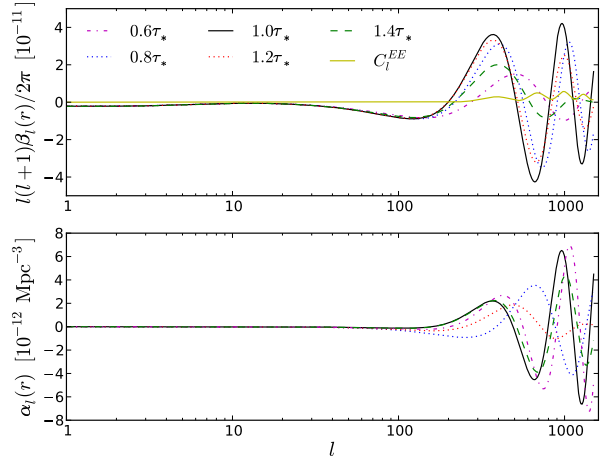}}}
\end{center}
\caption{Same as previous plot but for E-type polarisation.} 
\label{fig:abE}
\end{figure}

\subsubsection{The Three-to-One Power spectrum}

The number of power spectra that can be associated with a given
multispectrum depends on number of different ways the order of the
multispectrum can be decomposed into a pair of integers. The
bispectrum being of order three can be decomposed uniquely $3=2+1$
and has only one associated power spectrum. On the other hand the
order of the trispectrum can be decomposed in a two different ways,
i.e. $4=3+1=2+2$. Hence the trispectrum of a specific type generates
a pair of two different power spectrum associated with it; see, e.g.
\cite{Munshi_kurt} for more details. The results presented here are
generalizations for the case of non-zero spins.

The other power spectrum associated with the trispectrum is
constructed by cross-correlating product of three different fields
$[\calU\calV\calW]$ with the remaining field $[\calX]$. The
cross-correlation power spectrum in terms of the multipoles is given
by the following expression:

\beq
[\calU\calV\calW]_{lm} = \int [\calU(\oh)\calV(\oh)\calX(\oh)][{}_{u+v+w} Y_{lm}(\oh)^*]d\oh; \qquad
[\calZ]_{lm} = \int [\calX(\oh)][{}_{z} Y_{lm}(\oh)^*]d\oh; \qquad
C_l^{\calU\calV\calW,\calX} = {1 \over 2l + 1} \sum_{m=-l}^l [\calU\calV\calW]_{lm} [\calX]_{lm}^*.
\eeq

\n By repeated use of the expressions Eq.(\ref{eq:harmonics1}) or,
equivalently, Eq.(\ref{eq:harmonics2}) to simplify the harmonics of
the product field in terms of the individual harmonics, we can
express the all-sky result in the following form:

\beq
C_l^{\calU\calV\calW,\calX} = \sum_{l_1,l_2,l_3,L} T^{\calU_{l_1}\calV_{l_2}}_{\calW_{l_3}\calX_l}(L)
\left ( \begin{array}{ c c c }
     l_1 & l_2 & L \\
     u & v & -{(u+v)}
  \end{array} \right )
\left ( \begin{array} { c c c }
     L & l_3 & l' \\
     (u+v) & w & -{(u+v+w)}
  \end{array} \right )
\sqrt {  {(2l_1+1)(2l_2+1) \over ( 2L+1 )} }
\sqrt { { (2L+1)(2l_3+1) \over (2l+1) }}.
\eeq

\n In the case of of partial sky coverage with a generic mask
$w(\oh)$ the relevant expression for the Pseudo-$C_{\ell}$s will as
usual involve the harmonic transform of the mask.

\beqa
[\calU\calV\calW w]_{lm} = \int [\calU(\oh)\calV(\oh)\calW(\oh) w(\oh)][{}_{u+v+w} Y_{lm}^*(\oh)]d\oh; \qquad
[\calX w]_{lm} = \int [\calX(\oh)w(\oh)][{}_{z} Y_{lm}^*(\oh)]d\oh; \qquad \tilde C_l^{\calU\calV\calW,\calX} = M_{ll'}C_{l'}^{\calU\calV\calW,\calX}
\eeqa

\n The mixing matrix has the following expression in terms of
various spins involved and the power spectra of the mask introduced
above. Note that the mixing matrix is of different form compared to
what we obtained for two-to-one power spectra. This is related to
how various fields with different spins were combined to construct
these two estimators. In case of temperature trispectrum (spin-0)
the two mixing matrices take the same form.

\beq
M_{ll'}^{uvw,x} = {1 \over 4\pi }\sum_{l_a} (2l'+1)(2l_a+1) \left ( \begin{array}{ c c c }
     l & l_a & l' \\
     (u+v+w) & 0 & -{(u+v+w)}
  \end{array} \right )
\left ( \begin{array} { c c c }
     l & l_a & l' \\
     x & 0 & -x
  \end{array} \right ) |w_{l_a}|^2;
\eeq

The expressions derived here apply to a general mask and to
correlated Gaussian instrument noise; any non-Gaussianity from the
noise will have to be allowed for. These calculations lead to
optimal estimators, but the weights required  depend on the precise
model of non-Gaussianity being assumed. We present the models for
primordial non-Gaussianity in the next section and use them to
construct optimal estimator in our later discussions.

For a specific example we choose $\calU = \calV = \calW = \delta_T$
and $\cal X= E$. In this case, the three-to-one estimator takes the
following form:

\beqa
&& C_l^{TTT,E} = \sum_{l_1,l_2,l_3,L} T^{T_{l_1}T_{l_2}}_{T_{l_3}E_l}(L)
\left ( \begin{array}{ c c c }
     l_1 & l_2 & L \\
     0 & 0 & 0
  \end{array} \right )
\left ( \begin{array} { c c c }
     L & l_3 & l \\

     0 & 0 & 0
  \end{array} \right )
\sqrt {  {(2l_1+1)(2l_2+1) \over ( 2L+1 )} }
\sqrt { { (2L+1)(2l_3+1) \over (2l+1) }}; \nn \\
&& M_{ll'}^{000,2} = {1 \over 4\pi }\sum_{l_a} (2l'+1)(2l_a+1) \left ( \begin{array}{ c c c }
     l & l_a & l' \\
     0 & 0 &  0
  \end{array} \right )
\left ( \begin{array} { c c c }
     l & l_a & l' \\
     2 & 0 & -2
  \end{array} \right ) |w_{l_a}|^2.
\eeqa

The bispectrum is defined through a triangular configuration in the
multipole space,  the trispectrum with a quadrilateral (which can be
decomposed into two constituent triangles). The Wigner-$3j$  symbols
enforce these constraints at various levels for bispectrum and
trispectrum. The two-to-one power spectrum or skew-spectrum
introduced in previous sections essentially sums over all possible
configurations of the triangle obtained by keeping one of its sides
fixed. In an analogous manner the two probes of trispectrum
introduced here are linked with two different configuration in the
harmonic domain. The two-to-two power spectrum keeps the diagonal of
the quadrilateral fixed whereas the three-to-one power spectra keeps
one of the sides fixed while summing over all other possible
configurations.

It is possible to introduce a window optimized to search selectively
for information for a specific configuration for either the
bispectrum or the trispectrum. However such analysis which
introduces mode-mode coupling can not be generalized to the
arbitrary partial sky coverage as there is already a coupling of
modes because of non-uniform coverage of the sky.

Unlike the bispectrum, the trispectrum has a non-vanishing
contribution from the Gaussian component of a CMB map (in the same
way that, while the third moment of a univariate Gaussian about the
mean is zero,  the corresponding fourth-order moment is not). The
Gaussian contribution represents the unconnected component of the
total trispectrum which needs to be subtracted out. This can be done
by generating Monte-Carlo maps using an identical mask. The resulting
Gaussian part of the spectra can then be subtracted from the
estimates from the real data. The procedure is similar to the
analysis of temperature-only data; for more details see
\cite{Munshi_kurt}. Noise (assumed Gaussian) can also be subtracted
out following a similar procedure. The treatment requires a
hit-count map from a realistic scanning strategy, for non-uniform distribution of noise,
from where the variance of noise distribution in each pixel can be constructed.

To derive and simplify the above expressions we have used the
following results related to the overlap integral involving three
spin harmonics which generalizes the well-known {\it Gaunt
Integrals} involving spin-$0$ spherical harmonics.

\beq
\int {}_{s}Y_{lm}(\oh){}_{s'}Y_{l'm'}(\oh){}_{s''}Y_{l''m''}(\oh) d\oh =  \sqrt{(2l+1)(2l'+1)(2l''+1)\over 4\pi}
\left ( \begin{array}{ c c c }
     l & l' & l'' \\
     m & m' & m''
  \end{array} \right)
\left ( \begin{array}{ c c c }
     l & l' & l'' \\
     -s & -s' & -s''
  \end{array} \right).
\label{eq:harmonics1}
\eeq

\n The above result can be cast in following form which is useful
for expressing integral of more than three spin spherical harmonics
in terms of Wigner-$3j$ symbols:

\beq
{}_{s}Y_{lm}(\oh){}_{s'}Y_{lm}(\oh) = \sum_{LSM}{}_SY^*_{LM} \sqrt{(2L+1)(2l'+1)(2l''+1)\over 4\pi}
\left ( \begin{array}{ c c c }
     L & l' & l'' \\
     M & m' & m''
  \end{array} \right)
\left ( \begin{array}{ c c c }
     L & l' & l'' \\
     -S & -s' & -s''
  \end{array} \right).
\label{eq:harmonics2}
\eeq

\n The Wigner-$3j$ symbols satisfy an orthogonality relationship
which can be written as:

\beq
\sum_{m_1m_2} \left ( \begin{array}{ c c c }
     l_1 & l_2 & l_a \\
     m_1 & m_2 & m_a
  \end{array} \right)
\left ( \begin{array}{ c c c }
     l_1 & l_2 & l_b \\
     m_1 & m_2 & m_b
  \end{array} \right) = {\delta_{l_al_b}\delta_{m_am_b} \over 2l_a + 1}.
\eeq

\n In the remainder of the paper we will generalize the estimators
by including inverse variance weighting. We will first show how to
include the effect of noise and mask by using Monte-Carlo
simulations. Next we will consider the exact inverse variance
weighting where the inverse covariance matrix incorporates the
effect of noise and mask.

Note that incorporating a non-zero $B$-type (magnetic) polarisation
is possible in our formalism presented above by conceptually trivial
extension. We have only taken into account E-type (electric)
polarisation to keep the derivations simple as E-modes are assumed
to dominate presence of B-mode polarisation.

\section{The Bispectrum and Trispectrum with Primordial Non-Gaussianity}
\label{sec:cmb_bispec}

The optimization techniques that we will introduce in this section
follow the discussion in \cite{MuHe09} and \cite{Munshi_kurt}. The
optimization procedure depends on construction of three-dimensional
fields from the harmonic components of the temperature fields
$a_{lm}$ with suitable weighting with respective functions
$(\alpha,\beta,\mu)$ which describes primordial
non-Gaussianity\citep{YKW}. These weights make the estimators act in
an optimal way and the matched filtering technique adopted
ensures that the response to the observed non-Gaussianity is maximum
when it matches with the target primordial non-Gaussianity
corresponding to the weights.

In the linear regime, the curvature perturbations which generate the
fluctuations in the CMB sky are written as:

\begin{equation}
a^\calX_{lm} = 4\pi (-i)^l \int {d^3k \over (2\pi)^3} \Phi({\bf k}) \Delta_l^\calX(k)Y_{lm}(\hat k); \qquad \calX \in T,E.
\end{equation}
\n

\n We will need the following functions to construct the harmonic
space trispectrum as well as to generate weights for the
construction of optimal estimators:

\be
\alpha^\calX_l(r) = {2 \over \pi} \int_0^{\infty} k^2 dk \Delta^\calX_l(k)j_l(kr); ~~
\beta^\calX_l(r) = {2 \over \pi} \int_0^{\infty} k^2 dk P_{\phi}(k) \Delta^\calX_l(k)j_l(kr); ~~
\mu^\calX_l(r) = {2 \over \pi} \int_0^{\infty} k^2 dk \Delta^\calX_l(k)j_l(kr)
\ee

For a more complete description of the method for extracting the
trispectrum from a given inflationary model, see
\citet{Hu00},\cite{huOka02}.  In the limit of low multipoles, where the
perturbations are mainly dominated by Sachs-Wolfe Effect the
transfer functions $\Delta_l(k)$ take a rather simple form
$\Delta_l(k)=  j_l(kr_*)/3$ where $r_* = (\eta_0 - \eta_{dec})$
denotes the time elapsed between cosmic recombination and the
present epoch. In general the transfer function needs to be computed
numerically. The {\em local} model for the primordial bispectrum and
trispectrum can be constructed by going beyond linear theory in the
expansion of the $\Phi(x)$. Additional parameters $f_{NL}$ and
$g_{NL}$ are introduced which need to be estimated from observation.
As discussed in the introduction, $g_{NL}$ can be linked to $r$ the
{\em scalar to tensor ratio} in a specific inflationary model and
hence expected to be small. 

\begin{equation}
\Phi({\bf x}) = \Phi_L({\bf x}) + f_{NL} \left ( \Phi^2_L({\bf x}) - \langle \Phi_L^2({\bf x}) \rangle \right ) + g_{NL} \Phi_L^3({\bf x})
+ h_{NL} \left ( \Phi^4_L({\bf x}) - 3\langle \Phi_L^2({\bf x}) \rangle \right ) + \dots
\end{equation}

We will only consider the local model of primordial non-Gaussianity
in this paper and only adiabatic initial perturbations. More
complicated cases of primordial non-Gaussianity will be dealt with
elsewhere. In terms of the inflationary potential $V(\phi)$
associated with a scalar potential $\phi$ one can express these
constants as \citep{Hu00}:

\be
f_{\rm NL} = -{5 \over 6} {1 \over 8\pi G} {\partial^2 \ln V(\phi) \over \partial \phi^2}; \qtwo
g_{\rm NL} = {25 \over  54} {1 \over  (8 \pi G)^2} \left [ 2\left ( {\partial^2 \ln V(\phi) \over \partial \phi^2}\right )^2
- \left ( \partial^3 \ln V(\phi) \over \partial \phi^3 \right ) \left ( \partial \ln V(\phi) \over \partial \phi\right ) \right ].
\ee

There are two contributions to the primordial non-Gaussianity. The
first part is parametrized by $f_{\rm NL}$ and the second
contribution  is proportional to a new parameter which appears at
fourth order which we denote by $g_{\rm NL}$. From theoretical
considerations in generic models of inflation one would expect
$g_{\rm NL} \le r /50 $ \citep{Seery07}. Following \citet{Hu00} we
can expand the above expression in Fourier space to write:

\be
\Phi_2(k) = \int {d^3\bk_1 \over (2\pi)^3} \Phi_L(\bk+\bk_1)\Phi_L^*(\bk_1) -
(2\pi)^3 \delta_D(\bk)\int {d^3\bk_1 \over (2\pi)^3} P_{\Phi\Phi}(\bk_1);
~~~~~~\Phi_3(k) = \int {d^3\bk \over (2\pi)^3} \Phi_L(\bk_1)\Phi_l(\bk_2)\Phi_l^*(\bk_1).
\ee

\n The resulting trispectrum $T_{\Phi}$ for the potential $\Phi$
associated with these perturbations can be expressed as:

\begin{equation}
T_{\Phi}(k_1,k_2,k_3,k_4) \equiv \langle \Phi(\bk_1)\Phi(\bk_2)\Phi(\bk_3)\Phi(\bk_4)\rangle_c
= \int {d^3{\bK} \over (2 \pi)^3} \delta_D(\bk_1 +\bk_2 -\bK) \delta_D(\bk_3+\bk_4+\bK)
{\cal T}_{\Phi}(\bk_1,\bk_2,\bk_3,\bk_4,\bK);
\end{equation}

\n where the reduced trispectrum ${\cal
T}(\bk_1,\bk_2,\bk_3,\bk_4,\bK)$ can be decomposed into two distinct
contributions:

\be
{\cal T}_{\Phi}^{(2)}(\bk_1,\bk_2,\bk_3,\bk_4,\bK) = 4 f_{NL}^2 P_{\phi}(\bK) P_{\phi}(\bk_1) P_{\Phi}(\bk_3);~~~~
{\cal T}_{\Phi}^{(3)}(\bk_1,\bk_2,\bk_3,\bk_4,\bK) = g_{NL} \left \{ P_{\phi}(\bK) P_{\phi}(\bk_1) P_{\Phi}(\bk_3) +
P_{\phi}(\bK) P_{\phi}(\bk_1) P_{\Phi}(\bk_3).
\right \}
\ee

\n The resulting mixed trispectrum, involving temperature and
$E$-type polarization now can be written as:

\ben
T^{\calU_{l_1}\calV_{l_2}}_{\calW_{l_3}\calX_{l_4}}(L) =&& 4f_{\rm NL}^2 I_{l_1l_2L}I_{l_3l_4L}\int r_1^2dr_1 r_2^2 dr_2 F_L(r_1,r_2)
\alpha^\calU_{l_1}(r_1)\beta^\calV_{l_2}(r_1)\alpha^\calW_{l_3}(r_2)\beta^\calX_{l_4}(r_2)
\nonumber \\
&& + g_{\rm NL} I_{l_1l_2L}I_{l_3l_4L}\int r^2 dr \beta^\calV_{l_2}(r) \beta^\calX_{l_4}(r)
[ \mu^\calU_{l_1}(r)\beta^\calW_{l_3}(r) + \mu^\calW_{l_3}(r)\beta^\calU_{l_1}(r)].
\een

For detailed descriptions of objects such as $\alpha, \beta, \mu,
F_L$ see \citep{huOka02,Hu00,KomSpe01,Kogo06}. The CMB bispectrum
which describes departures from Gaussianity at the lowest level can
 be analysed in a similar fashion; see \cite{MuHe09} for a detailed
discussion and \cite{Smidt09} for a measurement in data. The corresponding expression is:

\be
B^{\calU\calV\calW}_{l_1l_2l_3} = 2f_{\rm NL} I_{l_1l_2l_3}\int r^2 dr \left [ \alpha^\calU_{l_1}(r) \beta^\calV_{l_2}(r)\beta^\calW_{l_3}(r) +
\alpha_{l_2}^\calV(r) \beta_{l_3}^\calW(r)\beta^\calU_{l_1}(r) + \alpha^\calW_{l_3}(r) \beta^\calV_{l_2}(r)\beta^\calU_{l_1}(r)  \right ].
\ee

The extension to orders beyond those presented here involves
higher-order Taylor coefficients and may not be practically useful
as detector noise and the cosmic variance may prohibit any
reasonable signal-to-noise. Numerical evaluations of the functions
$\alpha,\beta,\mu$ can be performed by using the publicly available
Boltzmann solvers such as CAMB\footnote{http://camb.info} or CMBFAST \footnote{http://www.cmbfast.org}.

A Gaussian fluctuation field has zero bispectrum. However, even a
Gaussian map will have non-zero trispectrum; this corresponds to the
unconnected part of the trispectrum. Its contribution can be
expressed in terms of the cross power spectra associated with
contributing fields:

\ben
G^{\calU_{l_1}\calV_{l_2}}_{\calW_{l_3}\calX_{l_4}}(L) =
(-1)^{(l_1+l_3)}\sqrt{(2l_1+1)(2l_3+1)} C_{l_1}^{\calU\calV}C_{l_3}^{\calW\calX}\delta_{l_1l_2}\delta_{l_3l_4}\delta_{L0}+
(2L+1) \left [(-1)^{l_1+l_2+L} C_{l_1}^{\calU\calW}C_{l_2}^{\calV\calX} \delta_{l_1l_3}\delta_{l_2l_4} + C_{l_1}^{\calU\calX}C_{l_2}^{\calV\calW} \delta_{l_1l_4}\delta_{l_2l_3} \right ]
\een

In the case of Gaussian maps, the trispectrum can be expressed
completely in terms of the relevant cross power -spectra
$C_{l}^{\calU\calV}$ of corresponding fields $\calU$ and $\calV$.
The estimators designed in later sections estimate the combined
skew- or kurt-spectra and the Gaussian contributions are subtracted
accordingly. The Gaussian maps that are used for Monte-Carlo
estimates of bias and scatter are constructed to have same power
spectrum as the non-Gaussian maps being analysed.

\section{Partial Sky Coverage and Inhomogeneous Noise}

In this section, we consider the inverse-variance weighting of the
data. We will consider the all-sky case first and then  introduce
the analytical results that can handle data in the presence of
partial sky coverage as well as correlated Gaussian noise. The
method developed here relies on Monte-Carlo simulations to model
observational artefacts. These estimators uses a weighted version of
square temperature - temperature (or two-to-one) angular power
spectrum. In a manner similar to its simpler version introduced in
the previous section, this power spectrum extracts information from
the bispectrum as a function of length of one side of the triangle
in harmonic space, while summing over all possible configuration
given by the change of other two sides of the triangle.

\begin{figure}
\begin{center}
{\epsfxsize=7. cm \epsfysize=7. cm {\epsfbox[22 433  301 725]{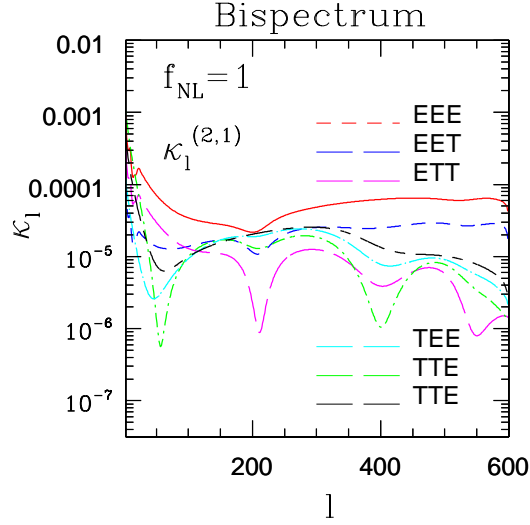}}}
\end{center}
\caption{Various bispectrum-related power spectrum $(2l+1){\cal K}^{(2,1)}_l$ plotted as function of angular scale $l$. We use $f_{NL}=1$ for each of these plots. Plots are based on WMAP7 \citep{WMAP7}
parameters equated out to lmax = 500.}
\label{fig:K31}
\end{figure}

\subsection{Bispectrum-related Power Spectrum or Skew-spectrum}

Following \citet{KSW}, we first construct the 3D
fields $A(r,\hat\Omega)$ and $B(r,\hat\Omega)$ from
the expansion coefficients of the observed CMB map, $a_{lm}$.
The harmonics here $A_{lm}(r)$ and $B_{lm}(r)$ are simply weighted
spherical harmonics of the temperature field $a_{lm}$ with weights constructed
from the CMB power spectrum $C_l$ and the functions $\alpha_l(r)$ and $\beta_l(r)$ respectively:
\begin{eqnarray}
A^\calU(r,\oh) \equiv \sum_{lm} Y_{lm}(\oh) A_{lm}^\calU(r);  ~~~ A_{lm}^\calU(r) \equiv \alpha_{l}^\calU(r)\sum_{\calU'} \sum_{l'm'}[{\bf C}^{-1}]_{\calU\calU'} b_{l'} a_{l'm'}^{\calU'}  = \alpha_l^\calU b_l \tilde a_{lm}^\calU; \\
B^\calU(r,\oh) \equiv \sum_{lm} Y_{lm}(\oh) B_{lm}^\calU(r); ~~~ B_{lm}^\calU(r) \equiv \beta_{l}^\calU(r) \sum_{\calU'} \sum_{l'm'} [{\bf C}^{-1}]_{\calU\calU'}b_{l'}a_{l'm'}^{\calU'} = \beta_l^\calU b_l \tilde a_{lm}^\calU.
\end{eqnarray}
\n The function $b_l$ represents beam smoothing, and from here
onward we will absorb it into the harmonic transforms.  The matrix
${\bf C}$ depends on both temperature (T) and $E$-type polarization
power spectra $C_l^{TT}$ and $C_l^{EE}$. The cross-correlation power
spectrum is denoted by $C_l^X$.

\beq
[{\bf C}]_l = \left ( \begin{array}{ c c }
     \calC_l^{TT} & \calC_l^{TE}  \\
     \calC_l^{TE} & \calC_l^{EE}
  \end{array} \right); ~~~~
{[\bf C]}^{-1}_l = {1 \over D_l} \left ( \begin{array}{ c c }
     -\calC_l^{EE} & \calC_l^{TE}  \\
     \calC_l^{TE} & -\calC_l^{TT}
  \end{array} \right);  ~~~~  \tilde a^\calU = [{\bf C}]^{-1}_{\calU\calU'} a^{\calU'}.
\label{eq:inv_cov}
\eeq

The determinant ${\cal D}_l$ introduced above
is a function of the relevant three power spectra introduced above ${\cal D}_l = 
C_l^{TT} C_l^{EE} - (C_l^{TE})^2$ for joint $(T,E)$ analysis. Using these definitions \citet{KSW} define the one-point mixed-skewness involving the fields $A^i(r,\hat \Omega)$ and $B^j(r,\hat \Omega)$ ($i,j,k \in {T,E}$):
\begin{equation}
S^{A^\calU B^\calV B^\calW} \equiv \int r^2 dr \int d \hat \Omega  A^\calU(r,\oh )B^\calV(r,\oh)B^\calW(r,\oh).
\end{equation}
\n $S^{A^\calU B^\calV B^\calW}$ can be used to estimate
$f^{loc}_{NL}$, but such a radical  compression of the data into a
single number restricts the  ability to estimate contamination of
the estimator by other sources of non-Gaussianity.  As a
consequence, we construct a less radical compression, to a function
of $l$, which can be used to estimate $f_{NL}^{\rm loc}$, but which
can also be analyzed for contamination by, for example, foregrounds.
We do this by constructing the integrated cross-power spectrum of
the maps $A^\calU(r,\hat\Omega)$ and
$B^\calV(r,\hat\Omega)B^\calW(r,\oh)$. Expanding $B^2$ in spherical
harmonics gives
\beq
[B^\calV{(\oh,r)}(r)B^\calW{(\oh,r)}]_{lm} \equiv \int d\hat \Omega B^\calV(r,\oh)B^\calW(r,\oh) Y_{lm}(\oh)
= \sum_{l'm'} \sum_{l''m''} {\beta_{l'}^\calV(r)}{\beta_{l''}^\calW(r)}
\left ( \begin{array}{ c c c }
     l & l' & l'' \\
     m & m' & m''
  \end{array} \right) I_{ll'l''}\tilde a^\calV_{l'm'} \tilde a^\calW_{l''m''}
\eeq
\n
and we define the cross-power spectrum $C_l^{A,B^2}(r)$ at a radial
distance $r$ as

\begin{equation}
C_l^{A^\calU,B^\calV B^\calW}(r) = {1 \over 2l + 1} \sum_m {\rm Real}\left \{ A^\calU_{lm}(r) [B^{\calV}(r)B^\calW(r)]_{lm} \right \},
\end{equation}
Integrating over $r$ we find:
\begin{equation}
C_l^{A^\calU,B^\calV B^\calW}\equiv \int r^2 dr~ C_l^{A^\calU,B^\calV B^\calW}(r).
\end{equation}
\n This integrated cross-power-spectrum of $B^2(r,\hat\Omega)$ and
$A(r,\hat\Omega)$ carries information about the underlying
bispectrum $B_{ll'l''}$, as follows:
\begin{eqnarray}
\hat C_l^{A^\calU,B^\calV B^\calW} =  {1 \over 2l + 1} \sum_m \sum_{l'm'}\sum_{l''m''}I_{ll'l'' }\left ( \begin{array}{ c c c }
     l & l' & l'' \\
     m & m' & m''
  \end{array} \right) \int r^2 dr \left \{ {\alpha_{l}^\calU(r)}{\beta_{l'}^\calV(r)}{\beta_{l''}^\calW(r)} \right \} \tilde a^\calU_{lm}
\tilde a^\calV_{l'm'} \tilde a^\calW_{l''m''}.
\end{eqnarray}
\n Similarly we can construct the cross-power-spectrum of the
product map $A^\calU B^\calV(r,\hat\Omega)$ and
$B^\calW(r,\hat\Omega)$, which we denote as
$C_l^{A^\calU B^\calV,B^\calW}$;
\begin{eqnarray}
\hat C_l^{A^\calU B^\calV,B^\calW} =  {1 \over 2l + 1} \sum_m \sum_{l'm'}\sum_{l''m''}
I_{ll'l''}
\left ( \begin{array}{ c c c }
     l & l' & l'' \\
     m & m' & m''
  \end{array} \right)
\int r^2 dr \left \{ {\beta^\calW_{l}(r)}
{\alpha_{l'}^\calU (r)}{\beta_{l''}^\calV (r)}  \right \}\tilde a^\calU_{lm}\tilde a^\calV_{l'm'}\tilde a^\calW_{l''m''}.
\end{eqnarray}
\n
Using these expressions, and the following relation, we can write this more compactly in terms of
the estimated CMB bispectrum
\begin{equation}
\hat B^{\calU~\calV~\calW}_{ll'l''} = \sum_{mm'm''} \left ( \begin{array}{ c c c }
     l & l' & l'' \\
     m & m' & m''
  \end{array} \right) a^\calU_{lm}a^\calV_{l'm'}a^\calW_{l''m''}
\end{equation}
\n from which we compute our new statistic, the {\em
bispectrum-related power spectrum}, $C_l^{\rm loc}$ as
\begin{equation}
\hat C_l^{A^\calU B^\calV B^\calW } \equiv (\hat C_l^{A^\calU,B^\calV B^\calW}
+  \hat C_l^{A^\calU B^\calV,B^\calW} + \hat C_l^{A^\calU B^\calW ,B^\calV}) =
{\hat f^{\rm loc}_{NL} \over (2l+1)} \sum_{\calU' \calV' \calW'}
\sum_{l'}\sum_{l''} \left \{ B^{\calU \calV \calW }_{ll'l''} [{\bf
C}^{-1}]^{\calU \calU'}_l [{\bf C}^{-1}]^{\calV \calV'}_{l'} [{\bf
C}^{-1}]^{\calW \calW'}_{l''} B^{\calU' \calV' \calW'}_{ll'l''}.
\right \} \label{BiPS}
\end{equation}

\n Clearly, in a joint analysis, the {\it pure} skew-spectrum such
as $C_l^{T,TT}$ or $C_l^{E,EE}$ discussed in the previous section
gets generalized to $C_l^{A^\calU, B^\calV B^\calW}$, etc, and the
construction of inverse covariance weighted fields mixes temperature
and $E$-type polarisation. Hence, each of these estimators carries
information from all possible types of mixed bispectra.  If we combine
various estimates into a unique power spectrum it compresses all
available information for $T$ and $E$-type polarisation maps:

\beq
\hat C_l^{(2,1)} = \sum_{\calU \calV \calW} \hat C_l^{A^\calU B^\calV B^\calW}
\eeq

\n where $B^{\rm loc}_{ll'l''}$ is the bispectrum for the local
$f_{NL}$ model, normalized to $f_{NL}^{\rm loc}=1$.  We can now use
standard statistical techniques to estimate $f_{NL}^{\rm loc}$. Note
that if we sum over all $l$ values then we recover the estimator
$S_{prim}$ of \citet{KSW}, which is the cross-skewness of $ABB$:

\begin{equation}
\hat S^{A_\calU B_\calV B_\calW }_3 = \sum_l (2l+1) ( \hat C_l^{A^\calU,B^\calV B^\calW} + \hat C_l^{A^\calU B^\calV,B^\calW} + \hat C_l^{A^\calU B^\calW,B^\calV} ).
\end{equation}

\n
If we sum over all possible triplets $ijk$, we can recover $S_3$ typically used in the literature and considered
previously \citep{MuHe09}.

\beq
\hat S_3 = \sum_{\calU \calV \calW} \hat S^{A_\calU B_\calV B_\calW}_3.
\eeq

\n Though $S_3$ compresses all available information at the level of
the bispectrum in temperature and polarization maps, it clearly also
has less power to distinguish any effect of systematics. These can
be studied in more detail if we carry out individual estimates which
break up the total into the estimates resulting from their various
linear combinations. The method develops here a simple extension of
previously used estimators both for one-point estimators as well as
two-point estimators or the associated power spectra. Such methods
will be useful with future surveys with higher signal to noise for
polarization measurements.

Recently, \citet{Cala09} studied the impact of secondaries in
estimation of primordial non-Gaussianity using temperature data.
Further studies by \citet{Hik09} used Fisher analysis to study joint
two-to-one analysis. The estimator introduced above shows how
individual linear combination of mixed bispectra can also be used
for joint estimation especially while dealing with non-primordial
contamination.

\subsection{Power Spectra related to the Trispectrum}

In a recent paper, \citet{Munshi_kurt} extended the earlier studies
by \citet{MuHe09} to the power spectrum related to the trispectrum in an
optimal way. This statistic was applied to WMAP 5-year data in \citep{Smidt2010}
to provide the first constraints on $\tau_{\rm NL}$ and $g_{\rm NL}$,
the third order corrections to primordial perturbations in
non-Gaussian models. The PCL estimators that we considered before are
generalizations of these estimators from the temperature-only case to
a fully joint temperature and polarization analysis. For a given
choice of mixed bispectrum, a pair of corresponding power spectra
can be defined which we optimize in this section for arbitrary
partial sky coverage and instrumental noise.

\subsubsection{Estimator for $\myK_l^{(3,1)}$}

\n Moving beyond the bispectrum, we can construct  estimators of the
two power spectra  we discussed above, $C_l^{(3,1)}$ and
$C_l^{(2,2)}$. We show how to decompose these entire estimators to
various choice of mixed bispectrum and compress the information
optimally to define an unique estimator for each power spectra. The
optimized versions for $\myK_l^{(3,1)}$ can be constructed by
cross-correlating the fields
$A^\calU(r_1,\oh)B^\calV(r_1,\oh)B^\calW(r_2,\oh)$ with
$B^\calX(r_2,\oh)$. In the first case the harmonics depend on two
radial distances ($r_1,r_2$) for any given angular direction. For a
specific combination of $\calU,\calV, \calW$ and $\calX$ which we
can choose either to be temperature $T$ or $E$-type polarization $E$
we can define a corresponding estimators. We will eventually combine
all various contributions that we recover from these combinations to
define a single estimator $\myK_l^{(3,1)}$ which will generalize the
estimator introduced in \citep{MuHe09} for analysis of
temperature-only data.

\be
A^\calU(r_1)B^\calV(r_1)|_{lm} = \int A^\calU(r_1,\oh)B^\calV(r_1,\oh)~Y_{lm}^*(\oh)~d\oh; \qquad
B(r_2)^\calX |_{lm} = \int B^\calX (r_2,\oh)~Y_{lm}^*(\oh)~d\oh.
\ee

\n Next, we construct the field $C^{\calU\calV}(r_1,r_2) = \sum_{lm}
F_{l}(r_1,r_2)A^\calU(r_1)B^\calV(r_1)|_{lm}Y_{lm}$. If we now form
the product of $C^{\calU\calV}(r_1,r_2)$ and $A^\calW(r_2)$ and
denote it as
$D^{\calU\calV\calW}(r_1,r_2)=C^{\calU\calV}(r_1,r_2)A^\calW(r_2)$
the spherical harmonic transform of this product field is
represented as $D_{lm}^{\calU\calV\calW}(r_1,r_2)$. Finally we
compute the  cross-power spectra between
$D^{\calU\calV\calW}(r_1,r_2)$ and $B^\calX(r_2)$. We denote it by
${\cal J}_l^{A^{\calU}B^{\calV}A^{\calW},B^\calX}(r_1,r_2)$ which
also depend on both radial distances $r_1$ and $r_2$:

\be
 {\cal J}_l^{A^\calU~B^\calV~A^\calW,B^\calX}(r_1,r_2) =
\frac{1}{2l+1} \sum_m {\re} \left [ \{ D^{\calU\calV\calW}(r_1,r_2) \}_{lm} \{ B^\calX(r_2)
\}_{lm}^* \right ].  \\
\ee

\n The construction for the second term is very similar. We start by
decomposing the real space product
$B^\calU(r,\oh)B^\calV(r,\oh)B^\calW(r,\oh)$ and $M^\calX(r,\oh)$ in
harmonic space. There is only one radial distance involved in both
of these terms.

\be
B^\calU(r,\oh)B^\calV(r,\oh)B^\calW(r,\oh)|_{lm} = \int [B^{\calU}(r,\oh)B^{\calV}(r,\oh)B^{\calW}(r,\oh)]~Y_{lm}^*(\oh)~d\oh; \qquad
M^\calX(r,\oh)|_{lm} = \int M^{\calX}(r,\oh) Y_{lm}^*(\oh)~d\oh.
\ee

\n Finally, the line-of-sight integral which involves two
overlapping contributions through the weighting kernels for the
first term and only one for the second gives us the required
estimator:

\be
{\cal K}_l^{(\calU\calV\calW,\calX)} = 4f_{\rm nl}^2 \int r_1^2 dr_1 \int r_2^2 dr_2  {\cal J}_l^{A^\calU~B^\calV~A^\calW,B^\calX}(r_1,r_2) +
2g_{\rm nl} \int r^2 dr {\cal L}_l^{B^\calU~B^\calV~B^\calW,M^\calX}(r).
\label{eq:k31}
\ee

Next we show that the construction described above does reduces to an optimum estimator for the power spectrum
associated with the trispectrum. The harmonics associated with the product field $B^\calU(r_1)B^\calV(r_1)B^\calW(r_2)$ can be
expressed in terms of various $\beta(r)$ functions:

\be
B^\calU(r_1)B^\calV(r_1)B^\calW(r_2)|_{lm} = \sum_{LM} (-1)^M \sum_{lm,l_im_i} \tilde a^\calU_{l_1m_1}\tilde a^\calV_{l_2m_2}\tilde a^\calW_{l_3m_3} ~
\alpha^\calU_{l_1}(r_1)\beta^\calV_{l_2}(r_1)\alpha^\calW_{l_3}(r_2) {\cal G}_{l_1l_2L}^{m_1m_2M}{\cal G}_{Ll_3l}^{Mm_3m}.  \\
\ee

\n
The cross-power spectra ${\cal J}_l^{ABA,B}(r_1,r_2)$ can be simplified in terms of the following expression:

\be
{\cal J}_l^{A^\calU~B^\calV~A^\calW,B^\calX}(r_1,r_2) =  \frac{1}{2l+1} \sum_{LM}(-1)^M \sum_m \left \{ F_{L}(r_1,r_2)
\alpha_{l_1}^\calU(r_1)\beta^\calV_{l_2}(r_1)\alpha^\calW_{l_2}(r_2)\beta^\calX_l(r_2) \right \} \langle \tilde a^\calU_{l_1m_1}\tilde a^\calV_{l_2m_2}
\tilde a^\calW_{l_3m_3}\tilde a^\calX_{lm} \rangle {\cal G}_{l_1l_2L}^{m_1m_2M}{\cal G}_{l_3lL}^{m_3mM}.
\ee

\n The Gaunt integral describing the integral involving three
spherical harmonics is defined as follows:

\be
{\cal G}_{l_1l_2l_3}^{m_1m_2m_3} = \sqrt{(2l_1+1)(2l_2+1)(2l_3+1)\over 4 \pi}
\left ( \begin{array}{ c c c }
     l_1 & l_2 & l_3 \\
     0 & 0 & 0
  \end{array} \right)\left ( \begin{array}{ c c c }
     l_1 & l_2 & l_3 \\
     m_1 & m_2 & m_3
  \end{array} \right).\\
\ee

\n The second terms can be treated in an analogous way and the
result takes the following form:

\be
{\cal L}_l^{B^{\calU}B^{\calV}M^{\calW},B^{\calX}}(r) =
\frac{1}{2l+1} \sum_{LM} (-1)^M \sum_m
\left \{ \beta_{l_1}^\calU(r)\beta_{l_2}^\calV(r) \mu_{l_3}^\calW(r)\beta_l^\calX(r) \right \}
\langle \tilde a^\calU_{l_1m_1}\tilde a^\calV_{l_2m_2}\tilde a^\calW_{l_3m_3}\tilde a^\calX_{lm} \rangle
{\cal G}_{l_1l_2L}^{m_1m_2M}{\cal G}_{l_3lL}^{m_3mM} .
\ee

\n Finally, when combined these terms as in Eq.(\ref{eq:k31}), we
recover the following expression:

\be
\hat {\cal K}^{(\calU\calV\calW,\calX)}_l = {1 \over 2l+1} \sum_{l_1l_2l_3} \sum_L {1 \over (2L+1)}{
[{\bf C}^{-1}]^{\calU\calU'}_{l_1}[{\bf C}^{-1}]^{\calV\calV'}_{l_2}[{\bf C}^{-1}]^{\calW\calW'}_{l_3}[{\bf C}^{-1}]^{\calX\calX'}_{l}}
 T_{\calW_{l_3}\calX_{l}}^{\calU_{l_1}\calV_{l_2}}[L]{\left [{\hat T}^{\calU'_{l_1}\calV'_{l_2}}_{\calW'_{l_3}\calX'_{l}}[L]
-{\hat G}^{\calU'_{l_1}\calV'_{l_2}}_{\calW'_{l_3}\calX'_{l}}[L] \right ]} .
\ee

\n We have subtracted the Gaussian component from the estimator in
the last step. This is done by simulating Gaussian maps in a
Monte-Carlo chain and using the same mask and the noise as the real
data, weighting functions used for real data are also used on the
Gaussian realisations. The ensemble average of the Gaussian
realisations are then subtracted from the estimates from the real
data.

We can sum over all possible mixed bispectra to construct the
following combined estimator:

\be
\hat {\cal K}^{(3,1)}_l = \sum_{\calU\calV\calW\calX}\hat {\cal K}^{(\calU\calV\calW,\calX)}_l.
\ee

\n The estimator ${\cal K}_l^{(\calU\calV\calW,\calX)}$ depends
linearly both on $f_{\rm NL}^2$ and $g_{\rm NL}$.  In principle, we
can use the estimate of $f_{\rm NL}$ from a bispectrum analysis as a
prior, or we can use the estimators ${\cal S}_l^{(2,1)}$,  ${\cal
K}_l^{(3,1)}$ and  ${\cal K}_l^{(3,1)}$ to put joint constraints on
$f_{NL}$ and $g_{NL}$.   The former is better from a signal-to-noise point of view \citep{Smidt2010}. Computational evaluation of either of the
power spectra clearly will be more involved as a double integral
corresponding to two radial directions needs to be evaluated. Given
the low signal-to-noise associated with these power spectra, binning
will be essential.

\subsubsection{Estimator for $\myK_l^{(\calU\calV,\calW\calX)}$}

\n In an analogous way the other power-spectra associated with the
trispectrum can be optimized by the following construction. We start
by taking the harmonic transform of the product field
$A(r,\oh)B(r,\oh)$ evaluated at the same line-of-sight distance $r$:

\be
A^\calU(r,\oh)B^\calV(r,\oh)|_{lm} = \int A^\calU(r)B^\calV(r)~Y_{lm}^*(\oh)~d\oh; \qtwo
B^\calW(r,\oh)M^\calX(r,\oh)|_{lm} = \int B^\calW(r)M^\calX(r)~Y_{lm}^*(\oh)~d\oh,
\ee

\n
and contract it with its counterpart at a different distance. The corresponding power spectrum
(which is a function of these two line-of-sight distances $r_1$ and $r_2$) has a first term

\be
{\cal J}_l^{A^\calU~B^\calV,A^\calW~B^\calX}(r_1,r_2) = \frac{1}{2l+1}\sum_m F_l(r_1,r_2) A^\calU(r_1,\oh)B^\calV(r_1,\oh)|_{lm}.
A(r_2,\oh)^\calW B(r_2,\oh)^\calX |_{lm}^*
\ee

\n
Similarly, the second part of the contribution can be constructed by cross-correlating the product of 3D fields
$B^\calU(\oh,r_1)B^\calV(\oh,r_1)$ against $B^\calW(\oh,r_2)M^\calX(\oh,r_2)$ evaluated at the same radial distance $r$

\be
{\cal L}_l^{B^\calU B^\calV,B^\calW M^\calX}(r) = \frac{1}{2l+1}\sum_m B^\calU~(r,\oh)B^\calV(r,\oh)|_{lm}B^\calW(r,\oh)M^\calX(r,\oh)|_{lm}^*.
\ee

\n
Finally, the estimator is constructed by integrating along the line of sight distances:

\be
\hat {\cal K}_l^{(\calU\calV,\calW\calX)} = 4f_{nl}^2 \int r_1^2 dr_1 \int r_2^2 dr_2 ~{\cal J}_l^{A^\calU B^\calV ,A^\calW B^\calX }(r_1,r_2) +
2g_{nl} \int r^2 dr ~{\cal L}_l^{B^\calU B^\calV ,B^\calW M^\calX }(r).
\label{eq:k22}
\ee

\n To see they do correspond to an optimum estimator  we use the
harmonic expansions and follow the same procedure outlined before:

\ben
&& {\cal L}_l^{B^\calU~B^\calV,B^\calW M^\calX}(r)  =  \frac{1}{2l+1}  \sum_m (-1)^m \sum_{l_im_i}
\left \{ \beta^\calU_{l_1}(r)\beta^\calV_{l_2}(r)\beta^\calW_{l_3}(r)
\mu_{l_4}^\calX(r) \right \} \langle \tilde a^\calU_{l_1m_1} \tilde a^\calV_{l_2m_2}\tilde a^\calW_{l_3m_3}\tilde a^\calX_{l_4m_4} \rangle {\cal G}_{l_1l_2l}^{m_1m_2m}{\cal G}_{l_3l_4l}^{m_3m_4m} .\\
&& {\cal J}_l^{A^\calU B^\calV,A^\calW B^\calX}(r_1,r_2)  =  \frac{1}{2l+1} \sum_{m} (-1)^m \sum_{l_im_i} \left \{ F_l(r_1,r_2)\alpha^\calU_{l_1}(r_1)
\beta^\calV_{l_2}(r_1)\alpha^\calW_{l_3}(r_2)\beta^\calX_{l_4}(r_2) \right \} \langle \tilde a^\calU_{l_1m_1}\tilde a^\calV_{l_2m_2}\tilde a^\calW_{l_3m_3}\tilde a^\calX_{l_4m_4} \rangle {\cal G}_{l_1l_2l}^{m_1m_2m}{\cal G}_{l_3l_4l}^{m_3m_4m} .
\een

\n Here we notice that ${\cal J}_l^{AB,AB}(r_1,r_2)$ is  invariant
under exchange of $r_1$ and $r_2$ but  ${\cal J}_l^{BB,BM}(r_1,r_2)$
is not. Finally, joining the various contributions to construct the
final estimator, as given in Eq.(\ref{eq:k22}), which involves a
line-of-sight integration:

\be
\hat {\cal K}_l^{(\calU\calV,\calW\calX)} = {1 \over 2l+1} \sum_{\calU'\calV'\calW'\calX'}\sum_{l_im_i} {1 \over (2l+1)}
[{{\bf C}^{-1}]^{\calU\calU'}_{l_1}[{\bf C}^{-1}]^{\calV\calV'}_{l_2}[{\bf C}^{-1}]^{\calW\calW'}_{l_3}[{\bf C}^{-1}]^{\calX\calX'}_{l_4}}
{  T^{\calU_{l_1}\calV_{l_2}}_{\calW_{l_3}\calX_{l_4}}(l)
[ {\hat T}^{\calU'_{l_1}\calV'_{l_2}}_{\calW'_{l_3}\calX'_{l_4}}(l) -  {\hat G}^{\calU_{l_1}\calV_{l_2}}_{\calW_{l_3}\calX_{l_4}}(l) ] },
\ee
and the combined estimator
\be
{\cal K}_{l}^{(2,2)} = \sum_{\calU \calV\calW \calX } {\cal K}_l^{(\calU\calV,\calW\calX)}.
\ee
\n
The prefactors associated with $f_{\rm NL}^2$ and $g_{\rm NL}$ are different in the linear combinations  ${\cal K}_l^{(2,2)}$ and ${\cal K}_l^{(3,1)}$, and hence even without using information
from third-order we can estimate both from fourth order alone.

\n Similarly, the one-point cumulant involving both temperature and
$E$-type polarization at fourth order can be written in terms of the
the mixed trispectra as follows:

\be
\hat {\cal K}_l^{(4)} = \sum_l {(2l+1)} \hat {\cal K}_l^{(2,2)} =
\sum_l (2l+1)\sum_{\calU \calV\calW \calX } \hat {\cal K}_l^{(\calU\calV,\calW\calX)}
= \sum_l {(2l+1)} \hat {\cal K}_l^{(3,1)} = \sum_l (2l+1)\sum_{\calU \calV\calW \calX } \hat {\cal K}_l^{(\calU\calV\calW,\calX)}.
\ee

\n It is also possible to carry out the sum over the harmonics $l$
without summing over the field types. In this case we recover
independent one point estimators associated with each type of mixed
trispectrum.

\be
\hat {\cal K}^{\calU\calV\calW\calX} = \sum_l (2l+1) \hat {\cal K}_l^{(\calU\calV,\calW\calX)} =
\sum_l (2l+1) \hat {\cal K}_l^{(\calU\calV\calW,\calX)}.
\ee

\n We will next consider the correction terms for these estimators
due to absence of absence of spherical symmetry - which may be
broken either because of mask or due to the presence of detector
noise in an arbitrary scanning strategy.

\subsection{Correction in the Absence of Spherical Symmetry}

It was pointed out in \citet{Babich,Crem06,Yadav08} that in the
presence of a partial sky coverage, e.g. due to the presence of a
mask or because of the galactic foregrounds and the bright point
sources, as well as, in the case of non-uniform noise, spherical
symmetry is destroyed. The estimators introduced above will then
have to be modified by adding terms which are linear in the observed
map. The corrective terms are incorporated using Monte-Carlo
techniques. A more general treatment which involves computationally
expensive inverse covariance weighting will be discussed later. The
treatment discussed here is nearly optimal though ignores mode-mode
coupling dominant at low $\ell$.

\subsubsection{Corrective terms for ~$C_l^{\calU,\calV\calW}$}

The (linear) corrective terms are constructed from correlating the
Monte-Carlo (MC) averaged $\langle A^\calU (r,\oh)~B^\calV
(r,\oh)\rangle_{sim}$ product maps with the input $B^\calW(r,\oh)$
map. The mask and the noise  that are used in constructing the
Monte-Carlo averaged product map are exactly same as the observed
maps and the ones derived from them such as $A$ or $B$.
\n Mode-mode coupling is important  at low angular modes, and we
consider the full case later, but for higher frequency modes, we can
approximate the linear correction to the local shape:
\beqa
\hat C_l^{A^\calU~B^\calV~B^\calW} = && {1 \over f_{sky}}\left \{ {\hat C_l^{A^\calU,B^\calV~B^\calW} -
C_l^{\langle A^\calU,B^\calV \rangle B^\calW} - C_l^{\langle A^\calU,B^\calW \rangle B^\calV} - C_l^{A^\calU, \langle B^\calV B^\calW \rangle}} \right \}
 \nn \\
&& + {1 \over f_{sky}}\left \{{\hat C_l^{A^\calU B^\calV ,B^\calW} -
C_l^{\langle A^\calU B^\calV \rangle, B^\calW} -C_l^{B^\calV \langle A^\calU , B^\calW \rangle}
- C_l^{A^\calU \langle B^\calV ,B^\calW \rangle}} \right \}  \nn \\
&& + {1 \over f_{sky}}\left \{{\hat C_l^{A^\calU B^\calW ,B^\calV} -
C_l^{\langle A^\calU B^\calW \rangle, B^\calV} -C_l^{B^\calW \langle
A^\calU , B^\calV \rangle} - C_l^{A^\calU \langle B^\calW ,B^\calV
\rangle}} \right \} \eeqa \n where $f_{sky}$ is the observed sky
fraction.

The $C_l$s such as $C_l^{\langle AB \rangle ,B}$ describe the
cross-power spectra associated with Monte-Carlo (MC) averaged
product maps $\langle A(n,r)B(n,r) \rangle$ constructed with the
same mask and the noise model as the the observed map $B$. Likewise,
the term $C_l^{A\langle B,B \rangle}$ denotes the average
cross-correlation computed from MC averaging, of the product map
constructed from the observed map $A(\Omega,r)$ multiplied with a MC
realization of map $B(\Omega,r)$ against the same MC realization
$B(\Omega,r)$.
\n
\citet{Crem06} showed via numerical analysis that the linear terms are
less important in the equilateral case than in the local model. The use of such Monte-Carlo maps to model
the effect of mask and noise greatly improves the speed compared to full bispectrum analysis.

The use of linear terms was found to greatly reduce the scatter  of
this estimator, thereby improving its optimality. The estimator was
used in \citet{YaWa08} also to compute the $f_{NL}$ from combined
$T$ and $E$ maps. The analysis presented above is approximate,
because it uses a crude $f_{sky}$ approximation to deconvolve the
estimated power spectrum to compare with analytical prediction. A
more accurate analysis should take into account the mode-mode coupling which can dominate at low $l$. 
The speed of this analysis depends on how fast we can generate
non-Gaussian maps.  The
general expression which includes the mode-mode coupling will be
presented in the next section. However it was found by
\citet{YaWa08} that removing low $l$s from the analysis can be
efficient way to bypass the mode-mode coupling. A complete numerical
treatment for the case of two-point statistics such as $C_l^{A,B^2}$
will be presented elsewhere.

\subsubsection{Corrective terms for the Estimator~ ${\cal K}_l^{\calU \calV \calW, \calX}$}

For the four-point terms, we need to subtract linear and quadratic terms:
\ben
&&\hat {\cal J}_l^{A_1^\calU B_1^\calV A_2^\calW ,A_2^\calX} =
{1 \over f_{sky}} \left [ \tilde { \cal J}_l^{A_1^\calU B_1^\calV A_2^\calW ,B_2^\calX} - {\cal I}^{\rm Lin}_l -
{\cal I}^{\rm Quad}_l \right ] \\
&& { \cal I}^{\rm Lin}_l= {1 \over f_{sky}} \Big [
{\cal J}_l^{\langle A_1^\calU B_1^\calV A_2^\calW \rangle, B_2^\calX}
+{\cal J}_l^{A_1^\calU \langle B_1^\calV A_2^\calW, B_2^\calX \rangle}
+ {\cal J}_l^{B_1^\calV \langle A_1^\calU A_2^\calX , B_2^\calW \rangle}
+ {\cal J}_l^{A_2^\calW \langle A_1^\calU B_1^\calV, B_2^\calX \rangle}
\Big ]\\
&& { \cal I}^{\rm Quad}_l = {1 \over f_{sky}} \Big [
{\cal J}_l^{\langle A_1^\calU B_1^\calV \rangle A_2^\calW, B_2^\calX}
+ {\cal J}_l^{\langle A_1^\calU A_2^\calW \rangle B_1^\calV, B_2^\calX}
+ {\cal J}_l^{\langle B_1^\calV A_2^\calW \rangle A_1^\calU, B_2^\calX}
+ {\cal J}_l^{ A_1^\calU B_1^\calV \langle A_2^\calW, B_2^\calX \rangle }
+ {\cal J}_l^{ A_1^\calU A_2^\calW \langle B_1^\calV, B_2^\calX \rangle}
+ {\cal J}_l^{ B_1^\calV A_2^\calW \langle A_1^\calU, B_2^\calX \rangle}
\Big ]. 
\een

\n
The expressions are similar for the other terms that depend on only one radial distance:

\ben
&&\hat {\cal L}_l^{B^\calU B^\calV M^\calW ,B^\calX } = {1 \over f_{sky}} \left [ \tilde { \cal L}_l^{B^\calU B^\calV B^\calW ,M^\calX} - {\cal I}^{Lin}_l -
{\cal I}^{\rm Quad}_l \right ] \\
&& { \cal I}^{\rm Lin}_l= {1 \over f_{sky}} \Big [
{\cal L}_l^{B^\calU \langle B^\calV M^\calW , B^\calX \rangle} +  {\cal L}_l^{B^\calV \langle B^\calU M^\calW , B^\calX \rangle}
+ {\cal L}_l^{B^\calW \langle B^\calU M^\calV , B^\calX \rangle} +{\cal L}_l^{ \langle B^\calU B^\calV M^\calW \rangle, B^\calX} \Big] \\
&& { \cal I}^{\rm Quad}_l= {1 \over f_{sky}} \Big [
{\cal L}_l^{B^\calU \langle B^\calV M^\calW \rangle, B^\calX}
+{\cal L}_l^{B^\calV \langle B^\calU M^\calW \rangle, B^\calX} + {\cal L}_l^{B^\calW \langle B^\calU M^\calV \rangle, B^\calX}
+ {\cal L}_l^{B^\calU B^\calV \langle M^\calW, B^\calX \rangle} + {\cal L}_l^{B^\calV  B^\calW \langle M^\calU, B^\calX \rangle}
+ {\cal L}_l^{B^\calU B^\calW \langle M^\calV, B^\calX \rangle}  \Big].
\een

\n
To simplify the presentation we have used the symbol $A^\calU(r_1,\oh)= A^\calU_1; A^\calU(r,\oh)=A^\calU$ and so on. Essentially we can see
that there are terms which are linear in the input harmonics and terms which are quadratic in the input
harmonics. The terms which are linear are also proportional to the bispectrum of the remaining 3D fields
which are being averaged. On the other hand the prefactors for quadratic terms are
3D correlation functions of the remaining two fields. Finally putting all of these expressions we can write:

\be
\tilde {\cal K}_l^{(\calU\calV\calW,\calX)} = 4f_{\rm NL}^2 \int r_1^2 dr_1 \int r_2^2 dr_2 \tilde {\cal J}_l^{A^\calU B^\calV A^\calW ,B^\calX}(r_1,r_2)  + 2g_{NL}\int r^2 dr \tilde {\cal L}_l^{B^\calU B^\calV M^\calW, B^\calX}(r). \\
\ee

\n
From a computational point of view clearly the overlap integral $F_{L}(r_1,r_2)$
will be expensive and may determine to what resolution ultimately these direct techniques can be
implemented. Use of these techniques directly involving Monte-Carlo numerical simulations will be
dealt with in a separate paper (Smidt et al. in preparation). To what extent the linear and quadratic
terms are important in each of these contributions can only be decided by testing against simulation.

\subsubsection{Corrective terms for the Estimator~ ${\cal K}_l^{\calU~\calV,\calW~\calX}$}

\n The unbiased estimator for the other estimator can be constructed
in a similar manner. As before there are terms which are quadratic
in input harmonics with a prefactor proportional to terms involving
cross-correlation or variance of various combinations of 3D fields
and there will be linear terms (linear in input harmonics) with a
prefactor proportional to bispectrum associated with various 3D
fields.

\ben
&& \hat {\cal J}_l^{A_1^\calU B_1^\calV, A_2^\calW B_2^\calX } =
{1\over f_{sky}} \Big [\tilde {\cal J}_l^{A_1^\calU B_1^\calV,A_2^\calW B_2^\calX} - I_l^{\rm Lin} - I_l^{\rm Quad}  \Big ]  \\
&&  {\cal I}^{\rm Lin} =  {1\over f_{sky}}\Big [ {\cal J}_l^{A_1^\calU B_1^\calV ,\langle A_2^\calW B_2^\calX  \rangle}
+{\cal J}_l^{\langle A_1^\calU B_1^\calV \rangle,A_2^\calW B_2^\calX }+
{\cal J}_l^{A_1^\calU \langle B_1^\calV ,B_2^\calW \rangle A_2^\calX } + {\cal J}_l^{B_1^\calV \langle A_1^\calU ,B_2^\calW \rangle A_2^\calX} +
{\cal J}_l^{B_1^\calV \langle A_1^\calU,A_2^\calW \rangle B_2^\calX} + {\cal J}_l^{A_1^\calU \langle B_1^\calV,A_2^\calW \rangle B_2^\calX} \Big ] \\
&& {\cal I}^{\rm Quad} = {1 \over f_{sky}}\Big [ {\cal J}_l^{A_1^\calU \langle B_1^\calV, A_2^\calW B_2^\calX \rangle}
+{\cal J}_l^{B_1^\calV \langle A_1^\calU, A_2^\calW B_2^\calX \rangle}
+ {\cal J}_l^{\langle A_1^\calU B_1^\calV, A_2^\calW \rangle B_2^\calX} +{\cal J}_l^{\langle A_1^\calU B_1^\calV, B_2^\calX \rangle A_2^\calV}
\Big ] .
\een

\n
The terms such as  ${\cal K}_l^{AB,BM}(r_1,r_2)$ can be constructed in a very similar way. We display
the term  ${\cal K}_l^{AB^2,A}(r_1,r_2)$ with all its corrections included.

\ben
&& \hat {\cal L}_l^{B^\calU B^\calV,B^\calW M^\calX} = {1\over f_{sky}} \Big [\tilde {\cal L}_l^{B^\calU B^\calV ,B^\calW M^\calX } -
I_l^{\rm Lin} - I_l^{\rm Quad}  \Big ]  \\
&& {I}_l^{\rm Quad} = {1 \over f_{sky}}\Big [ {\cal L}_l^{B^\calU \langle B^\calV B^\calW\rangle,M^\calX} +
{\cal L}_l^{B^\calV \langle B^\calU,B^\calW \rangle M^\calX}
+ {\cal L}_l^{B^\calW \langle B^\calV,B^\calU \rangle M^\calX} +{\cal K}_l^{B^\calU B^\calV \langle B^\calW,M^\calX \rangle}
+{\cal K}_l^{B^\calW B^\calV \langle B^\calU,M^\calX \rangle}
+{\cal K}_l^{B^\calU B^\calW \langle B^\calV,M^\calX \rangle} \Big ] \\
&& {I}_l^{\rm Lin} = {1 \over f_{sky}}\Big [ {\cal J}_l^{B^\calU \langle B^\calV, B^\calW M^\calX \rangle} +{\cal J}_l^{B^\calV \langle B^\calU, B^\calW M^\calX \rangle}
+ {\cal J}_l^{\langle B^\calU B^\calV, M^\calX \rangle B^\calW} +{\cal J}_l^{\langle B^\calU B^\calV, B^\calW \rangle M^\calX} \Big ] .
\een

The importance of the linear terms  depends greatly on the target
model being considered. For example, while linear terms for
bispectral analysis can greatly reduce the amount of scatter in the
estimator for {\it local} non-Gaussianity, the linear term is less
important in modeling the {\it equilateral}  model. In any case the
use of such Monte-Carlo (MC) maps is known to reduce the scatter and
can greatly simplify the estimation of non-Gaussianity.  This can be
useful, as fully optimal analysis with inverse-variance weighting,
which treats mode-mode coupling completely, may only be possible on
low-resolution maps.

\be
\tilde {\cal K}_l^{(\calU\calV,\calW\calX)} = 4f_{\rm NL}^2 \int r_1^2 dr_1 \int r_2 dr_2 \tilde {\cal J}_l^{A^\calU B^\calV, A^\calW B^\calX}(r_1,r_2) 
 + 2g_{NL}\int r^2 dr \tilde {\cal L}_l^{B^\calU B^\calV, B^\calW M^\calX}(r). \\
\ee

\n
The corrections to one-point estimators can be recovered by performing appropriate sums. In the next
section we use direct summations and proper modelling of the covariance matrix as opposed to
the Monte-Carlo techniques used here. However in certain situation it may be difficult to model
the covariance matrix in an accurate way, we also provide analytical results which can handle
such situations.

\section{Exact Analysis of Optimal Estimators}
\label{sec:generalcase}

In the previous section we relied on Monte-Carlo simulations to
model the effect of finite sky coverage, noise as well as other
observational artefacts. The resulting method nearly optimal and for
most cases where mode mode coupling is not to strong i.e. for near
all-sky coverage can be efficient. To take mode-mode coupling
properly into account which is the case for low multipoles we need
to model the covariance matrix as accurately as we can. In this
section we tackle the case where inverse covariance weighting is
required. The resulting method is optimal and can provide accurate
results for studies with degraded resolution maps where
cross-contamination to primordial non-Gaussianities coming from
secondaries is minimum.

\subsection{The Power Spectrum Related to the Bispectrum}

\n The general expression for the bispectrum estimator was developed
by \citet{Babich} for arbitrary sky coverage and inhomogeneous
noise. The estimator includes a cubic term, which by
matched-filtering maximizes the response for a specific type of
input map bispectrum. The speed of this analysis depends on how fast
we can generate non-Gaussian maps. The linear terms vanish in the
absence of anisotropy but should be included for realistic noise to
reduce the scatter in the estimates; see Babich (2005) for details.
We define the optimal estimator as:
\begin{eqnarray}
&& \hat E_L^{\calX,\calY\calZ}[a] = \sum_{L'} [N^{-1}]_{LL'} \Big [ {1 \over 6} \sum_{\calX'\calY'\calZ'}\sum_{MM'} \sum_{ll'l_imm'm_i} B_{L'll'}^{\calX\calY\calZ}\left ( \begin{array}{ c c c }
     L' & l & l' \\
     M' & m & m'
  \end{array} \right) \nonumber \\ && ~~~~~~~~~~~\times \big \{ ([C^{-1}_{L'M',l_1m_1}]^{\calX\calX}a_{l_1m_1}^{\calX'})
([C^{-1}_{lm,l_2m_2}]^{\calY\calY'}a_{l_2m_2}^{\calY'})([C^{-1}_{l'm',l_3m_3}]^{\calZ\calZ'}a_{l_3m_3}^{\calZ'}) \nonumber \\
&& ~~~~~~~~~~- [C^{-1}_{lm,l'm'}]^{\calX\calY} ([C^{-1}_{L'M',l_2m_2}]^{\calZ\calZ'}a_{l_2m_2}^{\calZ'})-
 [C^{-1}_{LM,lm}]^{\calX\calZ} ([C^{-1}_{l'm',l_2m_2}]^{\calY\calY'}a_{l_2m_2}^{\calY'}) -
[C^{-1}_{LM,lm}]^{\calY\calZ} ([C^{-1}_{l'm',l_2m_2}]^{\calX\calX'}a_{l_2m_2}^{\calX'}) \big \} \Big ] ; \nonumber \\
&&  ~~~~~~~~~~~~~~~~~~~~~~~~~~~~~~~~~~~~~~~~~~~~~~~~~~~~~~~~~~~~~~~~~~~~~~~~~~~~~~~~~~~~~~~~ \calX,\calY,\calZ,\calX',\calY',\calZ' \in \{ T,E \}
\end{eqnarray}

\noindent where $N_{LL'}$ is a normalization to be discussed later.
A factor of ${1/(2l+1)}$ can be introduced with the sum $\sum_M$, if
we choose not to introduce the $N_{LL'}$ normalization constant.
This will make the estimator equivalent to the one introduced in the
previous section. As the data is weighted
 by $C^{-1}= (S+N)^{-1}$, or the inverse covariance matrix, the speed
of this analysis depends on how quickly we can generate non-Gaussian
maps. addition of higher modes will reduce the variance of the
estimator. In contrast, the performance of sub-optimal estimators
can degrade with resolution, due to the presence of inhomogeneous
noise or a galactic mask. However, an incorrect noise covariance
matrix can not only make the estimator sub-optimal but it will make
the estimator biased too. The noise model will depend on the
specific survey scan strategy. Numerical implementation of such
inverse-variance weighting or multiplication of a map by $C^{-1}$
can be carried out by conjugate gradient inversion techniques.
Taking clues from \citet{SmZa06}, we extend their estimators for the
case of the skew spectrum. We will be closely following their
notation whenever possible. First, we define $Q_L[a]$ and its
derivative $\partial_{lm}Q_L[a]$. The required input harmonics
$a_{lm}$ are denoted as $a$.
\begin{eqnarray}
&&\hat Q_L^{\calX,\calY\calZ}[a] \equiv {1 \over 6}  \sum_{M} a_{LM}^\calX \sum_{l'm',l''m''} B_{Ll'l''}^{\calX\calY\calZ}\left ( \begin{array}{ c c c }
     L & l' & l'' \\
     M & m' & m''
  \end{array} \right) a_{l'm'}^\calY a_{l''m''}^\calZ \\
&& \partial_{lm}^\calX \hat Q_L^{\calX,\calY\calZ}[a] \equiv {1 \over 6}\delta_{Ll} \sum_{l'm',l''m''} B_{Ll'l''}^{\calX\calY\calZ}\left ( \begin{array}{ c c c }
     L & l' & l'' \\
     m & m' & m''
  \end{array} \right) a_{l'm'}^\calY a_{l''m''}^\calZ; ~ \\
&& \partial_{lm}^{\calY/\calZ} \hat Q^{\calX,\calY\calZ}_L[a] \equiv
{1 \over 6 }\sum_{M}
 a^\calX_{LM}
\sum_{l'm'} B_{Lll'}^{\calX\calY\calZ}\left ( \begin{array}{ c c c }
     L & l & l' \\
     M & m & m'
  \end{array} \right) a^{\calZ/\calY}_{l'm'}.
\end{eqnarray}
\n These expressions differ from those for the one-point estimators
by the absence of an extra summation index. If summed over the free
index $L$, the expression for $Q_L$ reduces to a one-point
estimator. $Q_L^{\calX, \calY\calZ}[a]$ represents a map as well as
$\partial_{lm} Q^{\calX,\calY\calZ}_L[a]$, however
$Q^{\calX,\calY\calZ}_L[a]$ is cubic in input maps $a_{lm}$s where
as the derivatives  $\partial_{lm} Q^{\calX,\calY\calZ}_L[a]$ are
quadratic in input. The expression for the derivative, is different,
when the derivative is taken w.r.t the field (e.g. $\calX$ in this case)
associated with the free indices than when it is taken with respect
to a field $\calY$ or $\calZ$  whose indices are summed over.

The skew-spectrum can then be written as (the summation convention
is assumed for the next two equations):
\begin{equation}
\hat E_L^{\calX,\calY\calZ}[a] = [N^{-1}]_{LL'} \left \{ Q_{L'}^{\calX,\calY\calZ}[C^{-1}a] - \sum_{S} [C^{-1}a]_{lm}^S \langle \partial_{lm}^{S} Q_{L'}^{\calX,\calY\calZ}[C^{-1}a']\rangle_{MC}) \right \}; ~~~~~~~  S \in (\calX,\calY,\calZ)
\end{equation}

\n Here $\langle \rangle_{MC}$ denotes the Monte-Carlo averages. The
inverse covariance matrix in the harmonic domain
$[C^{XY}]^{-1}_{l_1m_1,l_2m_2}= \langle a_{l_1m_1}^\calX
a_{l_2m_2}^\calY \rangle^{-1}$ encodes the effects of noise and the
mask. For all-sky and in the signal-only limit, it reduces to the usual
$[C^{-1}]^{\calX\calY}_{l_1m_1,l_2m_2} = {1\over D^{\calX\calY}_l}
\delta_{ll'}\delta_{mm'}$ The normalization of the estimator which
ensures unit response can be written as:
\ben
&& N_{LL'} = {1 \over 6} \sum_{SS'} \Big [  \langle \Big \{ \partial^S_{l_1m_1}Q_L[C^{-1}a] \Big \}
[{\tilde C}_{l_1m_1,l_2m_2}]^{SS'}
\Big \{ \partial^{S'}_{l_2m_2}Q_{L'}[C^{-1}a] \Big \}  \rangle \nn \\
&& \qquad \qquad \qquad \qquad \qquad \qquad - \Big \{  \langle
\partial^S_{l_1m_1}Q_L[C^{-1}a]  \rangle \Big \}
[\calC^{-1}_{l_1m_1,l_2m_2}]^{SS'} \Big \{
\langle\partial^{S'}_{l_2m_2}Q_{L'}[C^{-1}a]\rangle \Big \} \Big ];
\qquad Q_L \equiv  Q_L^{\calX,\calY\calZ}; ~~~S,S' \in {\calX,\calY,
\calZ}.
\een 
\n In the above expression, and in those that follow,
we will not explicitly display the superscript on the normalization
matrix $N_{LL'}$ and $Q_L$ as they are obvious from the context. Summing
over repeated indices is assumed, and the second term ensures subtraction
of terms with self-coupling(s). We will be using the following
identity in our derivation:
\begin{equation}
\calC^{\calX\calY}_{l_1m_1,l_2m_2} \equiv \langle \tilde a^\calX_{l_1m_1}\tilde a^\calY_{l_2m_2}\rangle = \langle [C^{-1}a]^\calX_{l_1m_1} [C^{-1}a]^\calY_{l_2m_2} \rangle = \sum_{\calX'\calY'}\sum_{l_am_a}\sum_{l_bm_b} [C^{-1}]^{\calX\calX'}_{l_1m_1,l_am_a} C^{\calX'\calY'}_{l_am_a,l_bm_b} [C^{-1}]^{\calY\calY'}_{l_2m_2,l_bm_b}.
\end{equation}
\n
The Fisher matrix, encapsulating the errors and
covariances on the $E_L$, for a general survey associated
with a specific form of bispectrum can
finally be written as:
\begin{eqnarray}
F_{LL'}  = { 1 \over 36} \left ( {}_{(1)}\alpha^{PP}_{LL'} + {}_{(2)}\alpha^{PP}_{LL'} + {}_{(1)}\alpha^{QQ}_{LL'}+{}_{(2)}\alpha^{QQ}_{LL'}
+ {}_{(3)}\alpha^{QQ}_{LL'} + {}_{(4)}\alpha^{QQ}_{LL'} \right ).
%
\end{eqnarray}
\n Using the following expressions, which are extensions of
\citet{SmZa06}, we find that the Fisher matrix can be written as a sum
of two $\alpha$ terms $\alpha^{PP}$ and $\alpha^{QQ}$. The terms
involved $\alpha$ correspond to coupling only of modes that appear
in different $3j$ symbols. Self-couplings are represented by the
beta terms. The subscripts describes the coupling of various $l$ and
$L$ indices. The subscript $PP$ correspond to coupling of free
indices,i.e one free index $L_1$ with another free index $L_2$ and
similar coupling for indices that are summed over such as $l_1$,
$l_2$ etc. Similarly for subscript $QQ$ the free indices are coupled
with summed indices. Couplings are represented by the inverse
covariance matrices in the harmonic domain e.g. $C^{-1}_{lm,LM}$
denotes coupling of mode $LM$ with $lm$.
\begin{eqnarray}
&& {}_{(1)}\alpha^{PP}_{L_1L_2} =\sum_{M_1,M_2} \sum_{l_i l_i' m_i m_i'} B^{\calX\calY\calZ}_{ L_1l_1l_1'} B^{\calX\calY\calZ}_{L_2l_2l_2'}
\left ( \begin{array}{ c c c }
      L_1 & l_1 & l_1' \\
     M_1 & m_1 & m_1'
  \end{array} \right)\left ( \begin{array}{ c c c }
      L_2 & l_2 & l_2' \\
     M_2 & m_2 & m_2'
  \end{array} \right)
[\tilde \calC_{L_1M_1,L_2M_2}]^{\calX\calX}[\tilde \calC_{l_1m_1,l_2m_2}]^{\calY\calY}[\tilde \calC_{l_1'm_1',l_2'm_2'}]^{\calZ\calZ} \nn \\
&& {}_{(1)}\alpha^{QQ}_{L_1L_2} =\sum_{M_1,M_2} \sum_{l_il_i'm_im_i'}
 B^{\calX\calY\calZ}_{L_1l_1l_1'} B^{\calX\calY\calZ}_{L_2l_2l_2'}
\left ( \begin{array}{ c c c }
     L_1 & l_1 & l_1' \\
     M_1 & m_1 & m_1'
  \end{array} \right)\left ( \begin{array}{ c c c }
     L_2 & l_2 & l_2' \\
     M _2& m_2 & m_2'
  \end{array} \right)  [\tilde \calC_{L_1M_1,l_2m_2}]^{\calX\calY}[\tilde \calC_{l_1m_1,L_2M_2}]^{\calY\calX}[\tilde \calC_{l_1'm_1',l_2'm_2'}]^{ \calZ\calZ} \nn \\
&& \alpha^{PP}_{L_1L_2} = \wick[d]{123}{(<1 L_1 <2 l_1 <3 l_1')(>1 L_2 >2 l_2 >3 l_2')};~~~~~
\alpha^{QQ}_{L_1L_2} = \wick[d]{123}{(<1 L_1 <2 l_1 <3 l_1')(>2 L_2 >1 l_2 >3 l_2')}.
\end{eqnarray}
\n Similar results for ${}_{(2)}\alpha^{PP}_{L_1L_2}$ and
${}_{(2-4)}\alpha^{QQ}_{L_1L_2}$ can be obtained from permutative
reordering of the above results. The ordering of the multipole
indices and that of corresponding fields denoted by $\cal X$ is
important. For each different choice of field triplets we will have
a different set of skew-spectra associated with the bispectrum. If
we choose to have the same triplets for the primed and unprimed
fields then we will recover the Fisher matrix associated with that
specific choice. However, if we decide to choose a different set of
triplets then we will have the information about the level of
cross-contamination from one type of power spectra to another.
Within a specific choice of triplet e.g. $TEE$ choice to associate
the free index $L$ to a given field type e.g. $T$ or $E$ will
generate two different skew spectra from the same bispectrum
$B^{TEE}$.  Results  presented above will reduce to those of
\citet{SmZa06} when further summations over free indices $L_1$ and
$L_2$ are introduced to collapse the two-point object to the
corresponding one-point quantity. The $\beta$ terms that denote
cross-coupling are not presented here as they do not appear in the
final expressions for the Fisher matrix. A detailed analysis of
these terms is presented in \cite{MuHe09} and can be extended in a
very similar manner.
If we sum over $LL'$ the Fisher matrix reduces to a scalar $F=\sum_{LL'}F_{LL'}$
with, $\alpha^{PP}_{LL'} = \alpha^{QQ}_{LL'}= \alpha$ and   $\beta^{PP}_{LL'} = \beta^{PQ}_{LL'} =\beta^{QQ}_{LL'}= \beta$, where $\alpha$, $\beta$ and $F$ are exactly the same as introduced in \citet{SmZa06} for one-point estimators.

\subsubsection{Joint Estimation of multiple Bispectrum-related Power-Spectra}

\n The estimation technique described above can be generalized to
cover the bispectrum-related power spectrum associated with
different sets of bispectra (X,Y), where X and can Y can be one of the
combinations from the set $\{~(TTT), ~(TTE), ~(TEE), ~(EEE)
\}^{loc/eq}$. In addition to considering triplets corresponding to a
given primordial bispectrum, we can as well consider joint estimation
of different initial primordial bispectrum such as local ({\it loc})
bispectrum or equilateral ({\it eq}).
\begin{equation}
\hat E_L^{\Gamma}[a] =  \sum_S [F^{-1}]_{LL'}^{\Gamma\Gamma'} \left \{ Q_{L'}^{\Gamma'}[C^{-1}a] - \sum_{S'}[C^{-1}a]_{lm}^{S'} \langle \partial^{S'}_{lm} Q_{L'}^{\Gamma'}[C^{-1}a]\rangle_{MC}) \right \}; ~~~
\Gamma, \Gamma' \in {(TTT),(TTE),(TEE),(EEE)}^{loc/eq}; ~~S \in {T,E}.
\end{equation}
\n
The associated Fisher matrix now will consist of sectors $F_{LL'}^{\Gamma\Gamma}$,$F_{LL'}^{\Gamma\Gamma'}$ and
$F_{LL'}^{\Gamma'\Gamma'}$.
The sector $\Gamma\Gamma$ and $\Gamma'\Gamma'$ will in general will be related to errors associated with estimation of
bispectra of $\Gamma$ and $\Gamma'$ types, whereas the sector $\Gamma\Gamma'$ will correspond to their cross-correlation.
Clearly with a given mixed bispectra type it is possible to have different estimators by associating a
specific type of field $T$ or $E$ with the index which is not summed over.
\begin{equation}
F_{LL'}^{\Gamma\Gamma'}  = \left \{ {2\over 36} \left [ {}_1\alpha^{PP}_{LL'}
+\dots \right ]^{\Gamma\Gamma'} + {
4\over 36} [{}_1\alpha^{QQ}_{LL'} + \dots ]^{\Gamma\Gamma'} \right \}.
\end{equation}
\n Here the dots  represent contribution from other terms
represented by ${}_i\alpha$, which are obtained from permutation of
various indices. We have introduced the following notation above:
\begin{equation}
{}_1\left[\alpha^{PP}_{LL'} \right ]^{\Gamma\Gamma'} =
\sum_{MM'}
 \sum_{l_i l_i' m_i m_i'} B^{\Gamma}_{Ll_1l_1'} B^{\Gamma'}_{L'l_2l_2'} \left ( \begin{array}{ c c c }
     L & l_1 & l_1' \\
     M & m_1 & m_1'
  \end{array} \right)\left ( \begin{array}{ c c c }
     L' & l_2 & l_2' \\
     M' & m_2 & m_2'
  \end{array} \right)  [\tilde \calC_{LM,L'M'}]^{\calX\calX'}[\tilde \calC_{l_1m_1,l_2m_2}]^{\calY\calY'}[\tilde \calC_{l_1'm_1',l_2'm_2'}]^{\calZ\calZ'}\\
\end{equation}
\n and a similar expression holds for the other
$[\alpha^{PP}_{LL'}]$ terms as well as the $[\alpha^{QQ}_{LL'}]$
terms. We can also use the above formalism to obtain the
cross-correlation of a given skew-spectrum type from different kinds
of primordial bispectrum, as well as, say, local type and
equilateral types of non-Gaussianity.

\subsubsection{All-sky Homogeneous Noise}

The above expressions are very general results for arbitrary sky
coverage due to a specific scanning strategy. Our approach can deal
with complications resulting from partial sky coverage and inhomogeneous gaussian noise.
Any residual non-gaussian noise will have to be subtracted out and will need more elaborate
analysis of variance estimation.

If we now take the limiting case when we have all sky coverage and
homogeneous noise we can recover analytical results which are useful
for comparing various planned and ongoing surveys. In the the
all-sky limit, the covariance matrices are determined entirely by
signal and noise power spectra. To simplify the general expressions
derived so far for the case of all-sky coverage and uniform noise we
will use the following expression:

\begin{equation}
\calC^{\calX\calY}_{l_1m_1,l_2m_2} \equiv [C^{-1}]_{l_1} \delta_{l_1l_2}\delta_{m_1m_2} = \left ( \begin{array}{ c c c }
     {1 / d^{TT}_{l_1}} & {1 / d^{TE}_{l_1}}  \\
     {1 / d^{TE}_{l_1}} & {1 / d^{EE}_{l_1}}
  \end{array} \right) \delta_{l_1l_2}\delta_{m_1m_2}.
\end{equation}

\n
We recover following expression for the case of temperature:

\begin{equation}
F_{LL'}^{T,TT} =
{1 \over 36}  \left \{ 2 \delta_{LL'} \sum_{ll'}{{[B_{Lll'}^{TTT}]^2 \over d_L^T d_{l}^T d_{l'}^T} +
4\sum_{l} {[B_{LL'l}^{TTT}]^2 \over d_l^T d_{L}^T d_{L'}^T}} \right \}.
\end{equation}

\n If we assume that there is no correlation between the temperature
and $E$-type polarization then $d_l^T = C_l^T$ which reduces to the
temperature-only result. Notice that in this case the corresponding
${}_i\alpha^{PP}$ functions and ${}_i\alpha^{QQ}$ functions become
degenerate. Expressions for the case of estimation error of
$E_{LL'}^{T,EE}, E_{LL'}^{E,TE}$ can be obtained using the following
expressions for the related Fisher matrices:

\beqa
&& F_{LL'}^{T,EE} = {1 \over 36} \left \{ 2 \delta_{LL'} \sum_{ll'} { [B_{Lll'}^{TEE}]^2 \over d_L^T d_l^E d_{l'}^E } + 4 \sum_{l} { [B_{LL'l}^{TEE}]^2 \over d_L^X d_{L'}^X d_{l'}^E } \right \} \\
&& F_{LL'}^{E,TE} =  {1 \over 36} \left \{ \delta_{LL'}\sum_{ll'} [B_{Lll'}^{ETE}]^2 \left ( { 1 \over d_L^E d_{l}^T d_{l'}^E } +
 { 1 \over d_L^E d_{l}^X d_{l'}^X } \right ) +
\sum_{l}[B_{LL'l}^{ETE}]^2 \left ( {1 \over  d_L^X d_{L'}^X d_{l}^E} + {1 \over  d_L^E d_{L'}^E d_{l}^T}
+ {1 \over  d_L^X d_{L'}^E d_{l}^X} + {1 \over  d_L^E d_{L'}^X d_{l}^X} \right ) \right \}.
\eeqa

\n
The Fisher matrices for $F_{LL'}^{E,TT}$ and $F_{LL'}^{E,EE}$ can also be constructed in a similar manner.

Using a specific form for the bispectrum $b^{\rm loc}_{l_1l_2l_3}$,
such as the local model, the Fisher matrix elements can be further
expressed in terms of the transfer functions $\alpha^X$ and
$\beta^Y$ and the associated power spectra $C_{l}^{XY}$ using the
definitions of ${}_i\alpha^{PP}$ and ${}_i\alpha^{QQ}$ introduced
before:
\begin{eqnarray}
{}_1{[\alpha_{LL'}^{PP}]}^{\Gamma\Gamma} =
&& {f_{NL}^2 \over 4\pi} (2L+1)(2L'+1)\sum_{l}(2l+1)
\left ( \begin{array}{ c c c }
     L & L' & l \\
     0 & 0 & 0
  \end{array} \right)^2 { 1 \over d^\calX_{L}d^\calY_{L'}d^\calZ_l} \nonumber \\
&& \times \left \{ \int r^2 dr \left ( \alpha^\calX_{L}(r)\beta^\calY_{L'}(r)\beta^\calZ_l(r) +
\alpha^\calY_{L'}(r)\beta^\calX_{L}(r)\beta^\calZ_l(r) +
\alpha^\calZ_{l}(r)\beta^\calX_{L}(r)\beta^\calY_{L'}(r) \right )\right \}^2; \qquad \Gamma = \calX\calY\calZ     \\
{}_1{[\alpha_{LL'}^{QQ}]}^{\Gamma\Gamma} = && \delta_{LL'}~~
{f_{NL}^2 \over 4\pi}(2L+1) \sum_{ll'} (2l+1)(2l'+1)\left ( \begin{array}{ c c c }
     L & l & l' \\
     0 & 0 & 0
  \end{array} \right)^2 { 1 \over d^\calX_{L}d^\calY_{l}d^\calZ_{l'}} \nonumber \\
&& \times  \left \{\int r^2 dr \Big ( \alpha_{L}^\calX(r)\beta^\calY_{l}(r)\beta^\calZ_{l'}(r) +
\alpha^\calY_{l}(r)\beta^\calZ_{l'}(r)\beta^\calX_{L}(r) +
\alpha_{l'}^\calZ(r)\beta_{L}^\calX(r)\beta_{l}^\calY(r) \Big ) \right \}^2; \qquad \Gamma = \calX\calY\calZ.
\end{eqnarray}

\n The power spectra $C_l$ appearing in the denominator take
contributions from both the pure signal (i.e. the CMB) and the
detector noise. It is possible to bin the estimates in sufficiently
large bins that these are practically uncorrelated estimates for an
experiment such as Planck with very high sky-coverage. A joint
analysis will combine the the results from all possible estimators.
The cross-correlation among various estimators (characterized by
different choice of $\calX$,$\calY$ and $\calZ$) can be computed
following the same techniques.  A detailed analysis of the
singularity structure of the error-covariance matrix will be
presented elsewhere.

\subsection{Power Spectra related to Trispectra}

\n Extending our analysis to the case of four-point correlation
functions involving both temperature and polarization data, we will
next consider the case of one-point and two-point estimators which
are related to the mixed trispectra.

\subsubsection{One-point Estimators}

\n We will use an inverse-variance weighting for harmonics recovered
from the sky. The covariance matrix, expressed in the harmonic
domain, $\langle a_{lm}^\calU a_{l'm'}^\calV \rangle =
[C^{-1}]^{\calU\calV}_{lm,l'm'}$ is used to filter out modes
recovered directly from the sky $a_{lm}$. We use these harmonics to
construct optimal estimators. For all-sky coverage and homogeneous
noise, we can we recover the results derived in the previous section. We
start by keeping in mind that the trispectrum can be expressed in
terms of the harmonic transforms $a_{lm}$, which can either be
temperature multipoles or polarization multipoles :

\be
\tri = (2l+1) \sum_{m_i} \sum_M (-1)^M
\left ( \begin{array}{ c c c }
     l_a & l_b & L \\
     m_a & m_b & M
  \end{array} \right)
\left ( \begin{array}{ c c c }
     l_c & l_d & L \\
     m_c & m_d & -M
  \end{array} \right)
a^\calU_{l_am_a} \dots a^\calX_{l_dm_d} \qquad (i \in {a,b,c,d}), ~~(\calU,\calV, \calW, \calX) \in {T,E}.
\ee

\n Based on this expression we can devise a one-point estimator. In
the following discussion, the relevant harmonics can be based on
partial sky coverage.

\be
Q^{\calU\calV\calW\calX}[a] = {1 \over 4!} \sum_{LM} (-1)^M \sum_{l_im_i} \Delta(l_i;L) ~ \tri
 \left ( \begin{array}{ c c c }
     l_a & l_b & L \\
     m_a & m_b & M
  \end{array} \right)
\left ( \begin{array}{ c c c }
     l_c & l_d & L \\
     m_c & m_d & -M
  \end{array} \right)
a_{l_am_a}^\calU \dots a_{l_dm_d}^\calX.
\ee

\n The term $\Delta(l_i,L)$ is introduced here to avoid
contributions from Gaussian or disconnected contributions:
$\Delta(l_i,L)$ vanishes if any pair of $l_i$s becomes equal or
$L=0$ which effectively reduces the trispectra to a product of two
power spectra (i.e. disconnected Gaussian pieces). Its value is unity for the connected terms.
We will also need the first-order and second-order derivative with
respect to the input harmonics. The linear terms are proportional to
the first derivatives, and the quadratic terms are proportional to
second derivatives, of the function $Q[a]$, which is quartic in the
input harmonics.

\be
\partial_{lm}^\calU Q^{\calU\calV\calW\calX}[a] =  {1 \over 4!} \sum_{LM} (-1)^M  \sum_{l_im_i} \Delta(l_i;L) ~ \dtri
\left ( \begin{array}{ c c c }
     l & l_a & L \\
     m & m_a & M
  \end{array} \right)
\left ( \begin{array}{ c c c }
     l_b & l_c & L \\
     m_b & m_c & -M
  \end{array} \right)
a_{l_am_a}^\calV a_{l_am_b}^\calW a_{l_cm_c}^\calX.
\ee

\n The first-order derivative term such as $\partial^\calU_{lm}
Q^{(\calU\calV\calW\calX)}[a]$ is cubic in the input maps and the
second-order derivative is quadratic in input maps (in terms of
harmonics). However, unlike the estimator itself,
$Q^{(\calU\calV\calW\calX)}[a]$, which is simply a number, these
objects represent maps constructed from harmonics of the observed
maps. The quadratic terms contribute only to disconnected parts and
hence will not be considered. The optimal estimator for the
one-point cumulant can therefore be written as follows. This is
optimal in the presence of partial sky coverage and most general
inhomogeneous noise:

\be
E^{\calU\calV\calW\calX}[a] =  {1 \over N}  \left \{ Q^{\calU\calV\calW\calX}[C^{-1}a] -\sum_{S,lm} [C^{-1}a]^S_{lm} \langle \partial^S_{lm}Q^{\calU\calV\calW\calX}[C^{-1}a] \rangle \right \}; \qquad S \in (\calU,\calV,\calW,\calX).
\ee

\n The terms which are subtracted out can be linear or quadratic in
input harmonics. The linear term is similar to the one which is used
for bispectrum estimation, whereas the quadratic terms correspond to
the disconnected contributions and will vanish identically as we
have designed our estimators in such a way that it will not take any
contribution from disconnected Gaussian terms. The Fisher matrix
reduces to a number which we have used for normalization. The direct
summation over various harmonics as described above can be expensive
computationally and determines to what resolution the numerical
calculation can be performed. The input $[C^{-1}a]$ denote the
entire set of inverse variance weighted harmonics for the $Q$ and
$\partial Q$. It is interesting to note that modes after inverse
variance weighting are no longer pure and are linear combinations of
Temperature and $E$-type polarization modes. This true for the case
of estimators too. A given estimator $F^{\calU\calV\calW\calX}$
though correspond to a specific choice of trispectrum takes
contributions from modes which are themselves linear combinations of
pure mode types.

\ben
&& F^{\calU\calV\calW\calX}= {1 \over N^{}} = {\left ( 1 \over 4! \right )^2}
\sum_{\calU'\calV'\calW'\calX'} \sum_{LM} \sum_{L'M'} \sum_{(all~lm)} \sum_{(all~l'm')} (-1)^M (-1)^{M'}~\Delta(l_i;L)\Delta(l_i';L')~ \tri(L) \trip(L) \nn \\
&& \qquad \qquad  \times \left ( \begin{array}{ c c c }
     l_a & l_b & L \\
     m_a & m_b & M
  \end{array} \right)
\left ( \begin{array}{ c c c }
     l_c & l_d & L \\
     m_c & m_d & -M
  \end{array} \right)
\left ( \begin{array}{ c c c }
     l_a' & l_b' & L' \\
     m_a' & m_b' & M'
  \end{array} \right)
\left ( \begin{array}{ c c c }
     l_c' & l_d' & L' \\
     m_c' & m_d' & -M'
  \end{array} \right) \nn \\
&& \qquad \qquad  \qquad \qquad  \qquad \qquad  \qquad \qquad
\times \left \{ [\tilde {\cal C}_{l_am_a,l_a'm_a'}]^{\calU \calU'}\dots [\tilde {\cal C}_{l_dm_d,l_d'm_d'}]^{\calW \calW'} + {\rm cyc.perm} \right \}.
\een

\n The cyclic permutations in these terms will include
covariances involving all-possible permutations of the four fields
involved in construction of the mixed trispectrum and the related
power spectra. The ensemble average of this one-point estimator will
be a linear combination of parameters $f_{\rm NL}^2$ and $g_{\rm
NL}$. Estimators constructed at the level of three-point cumulants
\citep{SmZa06,MuHe09} can be used jointly with this estimator to put
independent constraints separately on  $f_{\rm NL}$ and $g_{\rm
NL}$. As discussed before, while the one-point estimator has the
advantage of higher signal-to-noise, such estimators are not immune
to contributions from an unknown component which may not have
cosmological origin, such as inadequate foreground separation. The
study of these power spectra associated with bispectra or trispectra
can be useful in this direction. Note that these direct estimators
are computationally expensive due to the inversion and
multiplication of large matrices, but can be implemented in
low-resolution studies where primordial signals may be less
contaminated by foreground contributions or secondaries.

Combining all possible choices of mixed trispectra it is possible to
introduce one single number which represents the entire information
content regarding the trispectrum from Temperature and $E$-type
polarization in the absence of $B$-modes. The corresponding
estimators and the Fisher matrix takes the following form:

\be 
E^{(4)} = \sum_{\calU\calV\calW\calX} E^{\calU\calV\calW\calX};
\qquad F^{(4)} =  \sum_{\calU\calV\calW\calX}
F^{\calU\calV\calW\calX}. 
\ee 
The choice of estimator is related to
the level of compromise one is willing to made to increase the
signal-to-noise at the expense of losing the crucial power to
distinguish contributions from various contributions. Clearly it is
also possible to design estimators by fixing a subset of all
available indices representing the choice of $T$ or $E$.

\subsubsection{Two-point Estimators}

Generalizing the above expressions for the case of the power
spectrum associated with trispectrum, we introduce two power-spectra
which we have discussed in previous section in the context of
construction of nearly-optimal estimators. The information content
in these power spectra are optimal, and when summed for over $L$ we
can recover the results of one-point estimators.

\be
Q^{(\calU\calV,\calW\calX)}_L[a] = {1 \over 4!} \sum_{M} (-1)^M \sum_{l_im_i} T^{\calU_{l_a}\calV_{l_b}}_{\calW_{l_c}\calX_{l_d}}(L)
 \left ( \begin{array}{ c c c }
     l_a & l_b & L \\
     m_a & m_b & M
  \end{array} \right)
\left ( \begin{array}{ c c c }
     l_c & l_d & L \\
     m_c & m_d & -M
  \end{array} \right)
a_{l_am_a}^\calU \dots a_{l_dm_d}^\calX.
\ee

\n The derivatives at first order and second order are as series  of
maps (for each $L$) constructed from the harmonics of the observed
sky. These are used in the construction of linear and quadratic
terms. We have retained the overall normalization factor $1 \over
4!$ so that our estimator reduces to the temperature-only estimator
introduced in \cite{Munshi_kurt}.

\be
\partial_{lm}^\calU Q^{(\calU\calV,\calW\calX)}_L[a] = {1 \over 4!} \sum_{T} (-1)^M \sum_{l_im_i}\Delta(l_i;L) T^{\calU_{l_a}\calV_{l_b}}_{\calW_{l_c}\calX_{l_d}}(L)
 \left ( \begin{array}{ c c c }
     l & l_a & L \\
     m & m_b & M
  \end{array} \right)
\left ( \begin{array}{ c c c }
     l_b & l_c & L \\
     m_b & m_c & -M
  \end{array} \right)
a_{l_am_a}^\calV \dots a_{l_cm_c}^\calX.
\ee

\n We can construct the other estimator in a similar manner. To
start with, we define the function
$Q^{(\calU,\calV\calW\calX)}_L[a]$ and construct its first and
second derivatives. These are eventually used for construction of
the estimator $E^{(\calU,\calV\calW\calX)}_L[a]$. As we have seen,
both of these estimators can be collapsed to a one-point estimator
$Q^{\calU\calV\calW\calX}[a]$. As before, the variable $a$ here
denotes input harmonics $a_{lm}^\calU$ recovered from the noisy
observed sky.

\be
Q^{(\calU,\calV\calW\calX)}_L[a] = {1 \over 4!} \sum_{M} \sum_{ST} (-1)^T  a_{LM}^\calU \sum_{l_im_i}
\Delta(l_i,L;T) T^{\calU_{L}\calV_{l_b}}_{\calW_{l_c}\calX_{l_d}}(T)
 \left ( \begin{array}{ c c c }
     L & l_b & S \\
     M & m_b & T
  \end{array} \right)
\left ( \begin{array}{ c c c }
     l_c & l_d & S \\
     m_c & m_d & -T
  \end{array} \right)
a_{l_bm_b}^\calV \dots a_{l_dm_d}^\calX.
\ee

\n The derivative terms will have two contributing terms
corresponding to the derivative w.r.t. the free index $\{LM\}$ and
the terms where indices are summed over e.g. $\{lm\}$, which is very
similar to the results for the bispectrum analysis with the
estimator $Q^{(\calX\calY,\calZ)}_L[a]$. One major difference that
needs to be taken into account is the subtraction of the Gaussian
contribution. The function $\Delta(l_i,L)$ takes into account of
this subtraction.

\ben
&&\partial_{lm}^\calU Q^{(\calU,\calV\calW\calX)}_L[a] =  \delta_{lL}\sum_{ST}  \sum_{l_im_i} \Delta(l_i,L;T)
T^{\calU_{L}\calV_{l_2}}_{\calW_{l_3}\calX_{l_4}}(T)
 \left ( \begin{array}{ c c c }
     l & l_b & S \\
     m & m_b & T
  \end{array} \right)
\left ( \begin{array}{ c c c }
     l_c & l_d & S \\
     m_c & m_d & -T
  \end{array} \right)
a^\calV_{l_bm_b}\dots a^\calW_{l_dm_d} \\
&& \partial_{lm}^\calX Q^{(\calU,\calV\calW\calX)}_L[a] =  \sum_M \sum_{l_im_i} a^\calU_{LM} \sum_T
 \Delta(l_i,L;T) T^{\calU_{L}\calV_{l_2}}_{\calW_{l_3}\calX_{l_4}}(S)
 \left ( \begin{array}{ c c c }
     L & l & S \\
     M & m & T
  \end{array} \right)
\left ( \begin{array}{ c c c }
     l_b & l_c & S \\
     m_b & m_c & -T
  \end{array} \right)
a^\calV_{l_bm_b}a^\calW_{l_cm_c}.
\een

\n Using these derivatives  we can construct the estimators
$E_L^{(\calU,\calV\calW\calX)}$ and $E_L^{(\calU\calV,\calW\calX)}$:

\ben
E_L^{(\calU,\calV\calW\calX)} = N_{LL'}^{-1} \left \{ Q^{(\calU,\calV\calW\calX)}_{L'}[C^{-1}a] -\sum_S [C^{-1}a]^{S}_{lm} \langle \partial^S_{lm}Q^{(\calU,\calV\calW\calX)}_{L'}[C^{-1}a] \rangle \right \}; \qquad S \in {\calU,\calV,\calW,\calX} \\
E_L^{(\calU\calV,\calW\calX)} = N_{LL'}^{-1} \left \{ Q^{(\calU\calV,\calW\calX)}_{L'}[C^{-1}a] -\sum_{S}[C^{-1}a]^{S}_{lm} \langle \partial^S_{lm}Q^{(\calU\calV,\calW\calX)}_{L'}[C^{-1}a] \rangle \right \}; \qquad S \in {\calU,\calV,\calW,\calX}.
\een

\n where summation over $L'$ is implied.  The quadratic terms will
vanish,  as they contribute only to the disconnected part. The
normalization constants are the Fisher matrix elements $F_{LL'}$
which can be expressed in terms of the target trispectrum
$T^{l_1l_2}_{l_3l_4}(L)$ and inverse covariance matrices $C^{-1}$
used for the construction of these estimators. The Fisher matrix for
the estimator $E_L^{(\calU,\calV\calW\calX)}$, i.e.
$F_{LL'}^{(\calU,\calV\calW\calX)}$ can be expressed as:

\ben
&& [N^{-1}]_{LL'} = F_{LL'}^{(\calU,\calV\calW\calX)} =  \left ( {1 \over 4!} \right )^2 \sum_{ST,S'T'}\sum_{(all~lm,l'm')}(-1)^M (-1)^{M'} [\tri(S)] [\trip(S')] \nonumber \\
&& \qtwo \Delta(l_il;L)\Delta(l_i';L') \times \left ( \begin{array}{ c c c }
     l_a & l_b & S\\
     m_a & m_b & T
  \end{array} \right)
\left ( \begin{array}{ c c c }
     L & l_d & S \\
     M & m_d & -T
  \end{array} \right)
\left ( \begin{array}{ c c c }
     l_a' & l_b' & S' \\
     m_a' & m_b' & T'
  \end{array} \right)
\left ( \begin{array}{ c c c }
     L' & l_d' & S' \\
     M' & m_d' & -T'
  \end{array} \right) \nonumber \\
&& \qtwo \times \Big \{ [{\tilde {\cal C}}_{LM,L'M'}]^{\calU\calU'}
[{\tilde {\cal C}}_{l_am_a,l_a'm_a'}]^{\calV\calV'}[{\tilde {\cal C}}_{l_bm_b,l_b'm_b'}]^{\calW\calW'}
[{\tilde {\cal C}}_{l_cm_c,l_c'm_c'}]^{\calX\calX'} + \dots \nn \\
&& \qquad \qquad \qquad \qquad \qquad \qquad  +  [{\tilde {\cal C}}^{\calU\calV'}_{LM,l_a'm_a'}]^{\calU\calV'}
[{\tilde {\cal C}}_{l_am_a,L'M'}]^{\calV\calU'}
[{\tilde {\cal C}}_{l_bm_b,l_b'm_b'}]^{\calW\calW'}
[{\tilde {\cal C}}_{l_cm_c,l_c'm_c'}]^{\calX\calX'} + \dots \Big \} .
\een

\n The first set of terms can be recovered from the first  term by
permuting the multipole indices while still keeping the coupling of
the free indices ${LL'}$ intact. Similarly the second set of terms
represented by $\dots$ can be recovered from the second terms but
considering only coupling between free indices and the one that are
summed over. There will be a total of six terms of first type and
eighteen of the second type. Similarly, for the other estimator
$E_L^{(\calU\calV,\calW\calX)}$, the Fisher matrix
$F_{LL'}^{(\calU\calV,\calW\calX)}$ can be written as a function of
the associated trispectrum and the covariance matrix of various
modes. For further simplification of these expressions we need to
make simplifying assumptions for a specific type of trispectra, see
\cite{MuHe09} for more details for such simplifications in the
bispectrum.

\beqa
&& [N^{-1}]_{LL'} = F_{LL'}^{(\calU\calV,\calW\calX)} = \left ( {1 \over 4!} \right )^2 \sum_{M} \sum_{M'} \sum_{(all~lm)} \sum_{(all~l'm')} (-1)^M (-1)^{M'} \tri(L)\trip(L')
\left ( \begin{array}{ c c c }
     l_a & l_b & L \\
     m_a & m_b & M
  \end{array} \right)
\left ( \begin{array}{ c c c }
     L & l_d & L \\
     M & m_d & -M
  \end{array} \right) \nonumber \\
&& \qquad \qquad \times \left ( \begin{array}{ c c c }
     l_a' & l_b' & L' \\
     m_a' & m_b' & M'
  \end{array} \right)
\left ( \begin{array}{ c c c }
     L' & l_d' & L' \\
     M' & m_d' & -M'
  \end{array} \right) \Delta(l_il;L)\Delta(l_i';L')~\left \{ [{\tilde {\cal C}}_{l_am_a,l_a'm_a'}]^{\calU\calU'}\dots
[{\tilde {\cal C}}_{l_dm_d,l_d'm_d'}]^{\calX\calX'} +{\rm cyc.perm.} \right \}
\eeqa

\n Knowledge of the sky coverage and the noise characteristics
resulting from a specific scanning strategy is needed for
modelling of  $[C^{-1}]^{\calU\calU'}_{l_dm_d,l_d'm_d'}$. We will
discuss the impact of inaccurate modelling of the covariance matrix
in the next section. The direct summation we have used for the
construction of the Fisher matrix may not be feasible except for low
resolution studies. However a hybrid method may be employed to
combine the estimates from low-resolution maps using the exact method
with estimates from higher resolution maps using other faster but sub-optimal
techniques described in previous section. In certain situations when
the data is noise-dominated further approximations can be made to
simplify the implementation; a more detailed discussion will be
presented elsewhere. It is possible to sum over all possible
trispectra to recover the entire information content:

\be
E_l^{(1,3)} = \sum_{\calU\calV\calW\calX}E_l^{(\calU,\calV\calW\calX)}; \qquad F_{LL'}^{(1,3)} =
\sum_{\calU\calV\calW\calX}F_{LL'}^{(\calU,\calV\calW\calX)}; \\
E_l^{(2,2)} = \sum_{\calU\calV,\calW\calX}E_l^{(\calU,\calV\calW\calX)}; \qquad F_{LL'}^{(1,3)} =
\sum_{\calU\calV\calW\calX}F_{LL'}^{(\calU,\calV\calW\calX)}.  \\
\ee

\n Summing over the free indices we recover the one-point estimators
and the corresponding Fisher matrices:

\be
E^{(4)} = \sum_l E_l^{(1,3)} =  \sum_l E_l^{(2,2)}; \qquad   F^{(4)}  = \sum_{LL'} F_{LL'}^{(2,2)} = \sum_{LL' }F_{LL'}^{(1,3)}.
\ee

\n
So far we have assumed that the covariance matrix can be modelled accurately. In the next section we will discuss the
impact of not knowing the covariance matrix accurately. We will show that though the estimators still remain
unbiased but they no longer remain optimal.

\begin{figure}
\begin{center}
{\epsfxsize=10. cm \epsfysize=10. cm {\epsfbox[22 144  590 714]{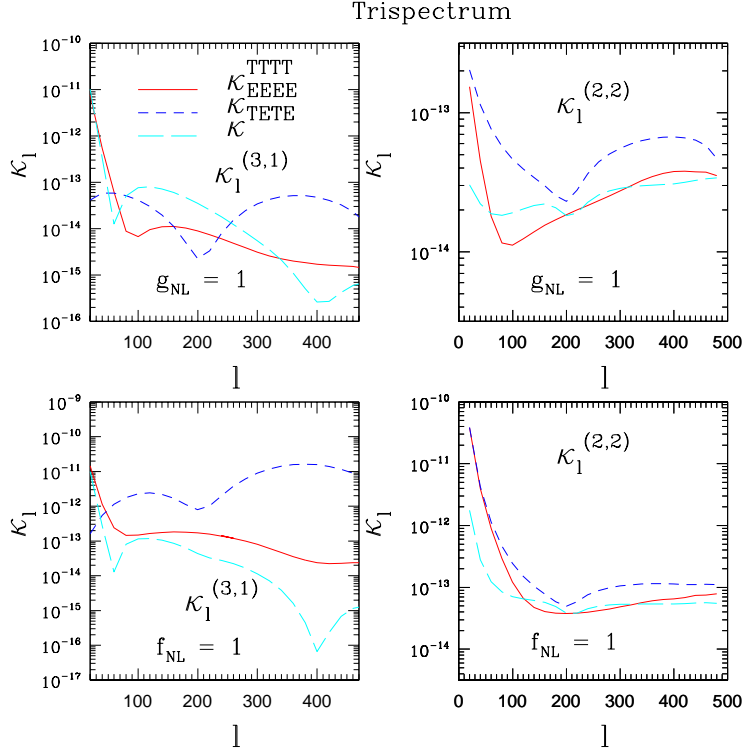}}}
\end{center}
\caption{Various Trispectrum related power-spectra are plotted as a function of
angular scale. An ideal all-sky no-noise experimental set up was used for 
computing the power spectra. Three different trispectra were considered for the
combinations $TTTT$, $EEEE$ and $TETE$. The left panels correspond to the estimator
$(2l+1){\cal K}_l^{(3,1)}$ and the right panels correspond to $(2l+1){\cal K}_l^{(2,2)}$.
Upper panels correspond to $g_{NL}=1$ and the bottom panels correspond to $f_{NL}=1$.
Cosmological parameters correspond to that of WMAP7 analysis \citep{WMAP7}. See text for details.}
\label{fig:Tris}
\end{figure}

\subsubsection{Approximation to exact $C^{-1}$ weighting and non-optimal weighting}

If the covariance matrix is not accurately known which is most often the case
due to the lack of exact beam or noise characteristics, as well as due to limitations on computer resources to model
it to high accuracy, it can be approximated. An approximation $R$ of $C^{-1}$ then acts as a regularization method.
The corresponding generic estimator can then be expressed as:

\be
\hat E_L^{Z}[a] = \sum_{L'} [F^{-1}]_{LL'} \left \{ Q_{L'}^Z[Ra]- \sum_S [Ra]_{lm}^S \langle \partial^S_{lm}Q^Z_{L'}[Ra] \rangle \right \};
\qtwo Z \in \{(\calU\calV,\calW\calX),(\calU,\calV\calW\calX)\}; ~~ S \in {\calU,\calV,\calW,\calX} .
\ee

\n As before we have assumed sums over repeated indices and $\langle
\cdot \rangle$ denote Monte-Carlo (MC) averages. As is evident from
the notation, the estimator above can be of type
$E_L^{(\calU,\calV\calW\calX)}$ or $E_L^{(\calU\calV,\calW\calX)}$.
For the collapsed case $E_L^{(4)}$ can also be handled in a very
similar manner.

\be
\hat E^Z[a] = \sum_L E_L^Z = \sum_{LL'}{ [F^{-1}]_{LL'}} \left \{ Q^Z_L[Ra]- [Ra]_{lm}^S \langle \partial^S_{lm}Q^Z_L[Ra] \rangle \right \}
\qtwo Z \in \{(\calU\calV,\calW\calX),(\calU,\calV\calW\calX)\}; ~~~~~S \in {\calU,\calV,\calW,\calX}.
\ee
We will drop the superscript $Z$ for simplicity, but any conclusion
drawn below will be valid for both specific cases i.e. $Z \in
\{(\calU\calV,\calW\calX),(\calU,\calV\calW\calX)\}$. The
normalization constant which acts also as inverse of associated
Fisher matrix $F_{LL'}$ can be written as:

\ben
&& F_{LL'} = \langle({\hat E_L})(\hat E_{L'})\rangle  -\langle({\hat E_L})\rangle \langle (\hat E_{L'})\rangle
 = {1 \over 4}  \sum_{SS'} \left \{ \langle \partial^S_{lm}
 Q_L[Ra] [C^{-1}]^{SS'}_{lm,l'm'} \partial^{S'}_{l'm'}Q_{L'}[Ra] \rangle - \langle \partial^{S}_{lm}Q_L[Ra]
\rangle [C^{-1}]^{SS'}_{lm,l'm'} \langle \partial^{S'}_{lm}Q_{L'}[Ra] \rangle \right \}; \nn \\
&& \qquad \qquad \qquad \qquad \qquad \qquad \qquad \qquad \qquad \qquad \qquad \qquad \qquad \qquad   S,S' \in {(\calU,\calV,\calW,\calX)}.
\een

\n
The construction of $F_{LL'}$ is equivalent to the calculation presented for the case of $R=C^{-1}$. For the
one-point estimator we similarly can write $F^R = \sum_{LL'} F_{LL'}^{R}$. The optimal weighting can be
replaced by arbitrary weighting. As a special case we can also use no weighting at all
$R = I$. This reduces the cost of the estimator drastically.  Although the estimator still remains unbiased
the scatter however increases as the estimator is no longer optimal. Use of arbitrary  weights makes the estimator
equivalent to a PCL estimator discussed before. In certain circumstances the use of a fast method can be
very useful before applying more robust and optimal techniques.

\subsubsection{Joint Estimation of Multiple Mixed Trispectra}

It may be of interest to estimate several trispectra jointly. The
different sources of trispectra can be of all primordial type such
as from ``adiabatic'' and ``isothermal'' perturbations. Such an
estimation can explore the joint error budget on parameters involved
from the same data-set. In such scenarios it is indeed important to
construct a joint Fisher matrix which will take the form

\be
\hat E_L^{\Gamma}[a] = \sum_{\Gamma\Gamma'} \sum_{LL'} [F^{-1}]^{\Gamma\Gamma'}_{LL'}
\hat E_{L'}^{\Gamma'}[a]; \qquad \qquad  {\Gamma,\Gamma' \in Adiabatic, Isothermal}.
\ee

\n The estimator $\hat E_L^{X}[a]$ is generic and it could be either
$E^{(3,1)}$ or $E^{(2,2)}$. Here $X$ and $Y$ corresponds to
different trispectra of type $X$ and $Y$, these could be e.g.
primordial trispectra from various inflationary scenarios. It is
possible of course to do a joint estimation of primary and secondary
trispectra. The off-diagonal blocks of the Fisher matrix will
correspond to the cross-talk between various types of bispectra.
Indeed, a  principal component  or generalized eigenmode analysis
can be useful in finding how many independent components of such
trispectra can be estimated from the data.

The cross terms in the Fisher matrix elements will be of following type:

\ben
&& F^{XY}_{LL'} =  \left ( {1 \over 4!} \right )^2 \sum_{ST,S'T'}\sum_{(all~lm,l'm')}(-1)^T (-1)^{T'}
\left [T^{\calU_{l_a}\calV_{l_b}}_{\calW_{l_c}\calX_{l_c}}(L) \right ]^\Gamma
\left [T^{l_{\calU_{l_a'}}\calV_{l_{b'}}}_{\calW_{l_{c'}}\calX_{l_{d'}}}(L') \right ]^{\Gamma'} \nonumber \\
&& \qquad  \times \left ( \begin{array}{ c c c }
     l_a & l_b & S\\
     m_a & m_b & T
  \end{array} \right)
\left ( \begin{array}{ c c c }
     L & l_d & S \\
     M & m_d & -T
  \end{array} \right)
\left ( \begin{array}{ c c c }
     l_a' & l_b' & S' \\
     m_a' & m_b' & T'
  \end{array} \right)
\left ( \begin{array}{ c c c }
     L' & l_d' & S' \\
     M' & m_d' & -T'
  \end{array} \right) \nonumber \\
&&  \qquad \qquad \qquad\qquad \qquad  \times \Big \{
[\tilde {\cal C}_{LM,L'M'}]^{\calU\calU'}[\tilde {\cal C}_{l_am_a,l_a'm_a'}]^{\calV\calV'}[\tilde {\cal C}_{l_bm_b,l_b'm_b'}]^{\calW\calW'}
[\tilde {\cal C}_{l_cm_c,l_c'm_c'}]^{\calX\calX'} + \dots  \nn \\
&& \qquad \qquad\qquad \qquad \qquad \qquad\qquad \qquad  + [\tilde {\cal C}_{LM,l_a'm_a'}]^{\calU\calV'}[\tilde {\cal C}_{l_am_a,L'M'}]^{\calV\calU'}
[\tilde {\cal C}_{l_bm_b,l_b'm_b'}]^{\calW\calW'}[\tilde {\cal C}_{l_cm_c,l_c'm_c'}]^{\calX\calX'} + \dots \Big \}.
\een

\n The expression displayed above is valid only for
$E^{(\calU\calV,\calW\calX)}$, exactly similar results holds for the
other estimator $E^{(\calU,\calV\calW\calX)}$. For $X=Y$ we recover
the results presented in previous section for independent estimates.
As before we recover the usual result for one-point estimator for
$Q^{4}$ from the Fisher matrix of $Q_L^{(3,1)}$ or $Q_L^{(2,2)}$,
with corresponding estimator modified accordingly.

\be
F^{\Gamma\Gamma'} = \sum_{\Gamma\Gamma'} F^{\Gamma\Gamma'}_{LL'};
\qtwo \hat E^{\Gamma}[a] = \sum_{\Gamma\Gamma'} [F^{-1}]^{\Gamma\Gamma'} \hat E^{\Gamma'}[a].
\ee

A joint estimation can provide clues to cross-contamination from different sources of trispectra.
It also provide information about the level of degeneracy involved in such estimates. It is also
possible to do joint estimation involving two different types of power spectra associated with trispectra
or to include even the power spectrum associated with the bispectrum. The results presented in this
section can be generalized to include such cases too.

\section{Conclusion}

The ongoing all-sky survey performed by the Planck satellite will
complete mapping the CMB sky in unprecedented detail, covering a
huge frequency range. It will provide high resolution temperature
and polarization maps which will provide the cosmological community
with the opportunity to constrain available theoretical models with
unprecedented accuracy. Recent detection of non-Gaussianity in WMAP
data has added momentum to non-Gaussianity studies.


The temperature and polarization power spectra carry the bulk of the
cosmological information, though many degenerate early universe
scenarios can lead to similar power spectra. Higher-order
multi-spectra can lift these degeneracies. The higher-order spectra
are the harmonic transforms of multi-point correlation functions,
which contain information that can be difficult to extract using
conventional techniques. This is related to their complicated
response to the inhomogeneous noise and partial sky coverage.
Analysis of the higher order polarization statistics is more
complicated not only by relatively low signal-to-noise ratio and
their spinorial nature but also because of uncertainties in modeling
the polarized foreground. A practical advance is to form collapsed
two-point statistics, constructed from higher-order correlations,
which can be extracted using conventional power-spectrum estimation
methods. If the polarization data can be analyzed and handled
properly it can help to further tighten the constraints on
parameters describing non-gaussianity that are achieved by analysis
of temperature data alone.


In a recent study, \citet{Munshi_kurt} studied and developed three
different types of estimators which can be employed to analyze these
power spectra associated with higher-order statistics such as
bispectrum or trispectrum for analysis of temeprature data. In this
work we include polarization on top of this to further tighten the
constraints. The analysis of polarization data makes the analysis
bit involved because of the spinorial nature of the data. We start
with the MASTER-based approach  \citep{Hiv02} which is typically
employed to estimate pseudo-$C_l$s from the masked sky in the
presence of noise These are unbiased estimators but the associated
variances and scatter can be estimated analytically with very few
simplifying assumptions. We extend these estimators to study
higher-order correlation functions associated mixed multispectra
such as bispectrum and trispectrum which involve both temperature
and polarization. Ignoring the contributions from B-polarization
makes these estimators much simpler. These estimator can be very
useful for testing simulation pipeline running Monte-Carlo chain in
a rather smaller time scale to spot spurious contributions from
foreground or other secondary sources. Generalizing the
temperature-only estimators we develop estimators for
$C_l^{(\calU\calV,\calW)}$ for the skew-spectrum (3-point) for
specific but arbitrary choice of $\calU,\calV,\calW,\calX$ as well
as $C_l^{(\calU,\calV\calW\calX)}$ and
$C_l^{(\calU\calV,\calW\calX)}$ which are power spectra of fields
related to the specific choice of mixed trispectrum, or
kurt-spectrum involving temperature and $E$-type polarization
fields.


For our next step, we generalized the estimators employed by
\cite{YKW,YaWa08,Yadav08} to study the kurt-spectrum. These methods
are computationally less expensive and can be implemented using a
Monte-Carlo pipeline which involve the generation of 3D maps from the
cut-sky harmonics using radial integrations of a target theoretical
model bispectrum along the line of sight. The Monte-Carlo generation
of 3D maps is the most computationally expensive part and dominates
the calculation. Including polarization field increases the
computational cost. The technique nevertheless has been used
extensively, as it remains highly parallelisable and is near optimal
in the presence of homogeneous noise and near all-sky coverage. The
corrective terms involve linear and quadratic contributions for the
lack of spherical symmetry due to the presence of inhomogeneous
noise and partial sky coverage. These terms can be computed using a
Monte-Carlo chain for joint temperature and polarization data, but including polarization data requires the ability to handle
further complications with an added level of sophistication. We also
showed that the  radial integral involved at the three-point
analysis needs to extended to incorporate a double-integral for the
mixed trispectrum. For every choice of the bi- or trispectrum
involving a specific combination of temperature and polarization
field, related power spectra can always be defined. These can help
as a diagnosis for cross-contamination from non-primordial sources
in different available frequency channels.


Though the estimators based on Monte-Carlo analysis based method are
very fast, they do not accurately take care of the mode-mode coupling, which are present at least at low resolution. We have
developed estimators, which are completely optimal even in the
presence of inhomogeneous noise and arbitrary sky coverage
(e.g. \cite{SmZa06}). These can handle mode-mode coupling more accurately.
Extending previous work by \cite{MuHe09} which concentrated only on
the skew-spectrum, \citet{Munshi_kurt} showed how to generalize it
to the case of power spectrum related to the trispectrum. In this
study we have included polarization in a completely general manner
both at the level of the bispectrum and trispectrum. This involves
finding a fast method to construct and invert the joint covariance
matrix $C_{lml'm'}$ in multipole space. In most practical
circumstances it is possible only to find an approximation to the
exact joint covariance matrix, and to cover this we present analysis
for an approximate matrix which can be used instead of $C^{-1}$.
This makes the method marginally suboptimal but it remains unbiased.
The four-point correlation function also takes contributions which
are purely Gaussian in nature. The subtraction of these
contributions is again simplified by the use of Gaussian Monte-Carlo
polarized maps with the same power spectrum. A joint Fisher analysis
is presented for the construction of the error covariance matrix,
allowing joint estimation of trispectra contributions from various
polarized sources, primaries or secondaries. Such a joint estimation
give us fundamental limits on how many  sources of non-Gaussianity
can be jointly estimated from a specific experimental set up which
scans the sky for temperature as well as for polarization.


At the level of the bispectrum, primordial non-Gaussianity can for
many models be described by a single parameter $f_{NL}$. The two
degenerate power spectra related to the trispectrum we have studied
at the 4-point level, require two parameters, typically $f_{NL}$ and
$g_{NL}$. Use of the two power spectra will enable us to put
separate constraints on $f_{NL}$ and $g_{NL}$ without using
information from lower-order analysis of the bispectrum, but they
can all be used in combination (see \citet{Smidt2010}). Clearly at even higher-order more
parameters will be needed to describe various parameters ($f_{NL}$,
$g_{NL}$, $h_{NL}$, \dots) which will all be essential in describing
degenerate sets of power spectra associated with multispectra at a
specific level. Previous studies concentrated only on temperature
data, including information from polarisation data can improve the
constraints. At present the polarization data is dominated by noise,
but surveys such as Planck will improve the signal to noise
available in polarization data. Future all-sky high sensitivity
polarization surveys too can further improve the situation and our
estimators will be be useful for analysis of such data. Numerical
implementation of our estimators will be reported elsewhere. We have
ignored presence of $B$-mode polarisation but our formalism can be
extended to take into account the magnetic or $B$-type polarisation.


The analytical results presented here can also be useful in the
context of study of shear data from weak lensing surveys. We plan to
present our results elsewhere (Munshi et al. 2010).

\section{Acknowledgement}

DM acknowledges support from STFC standard grant ST/G002231/1 at
School of Physics and Astronomy at Cardiff University. This work was
initiated when DM was supported by a STFC postdoctoral fellowship at
the Royal Observatory, Institute for Astronomy, Edinburgh. It is a
pleasure to thank Walter Gear for useful exchanges. We also
acknowledge many useful discussions with Michele Liguori.
AC and JS acknowledge support from NSF AST-0645427 and NASA NNX10AD42G.

\bibliography{pol.bbl}

\end{document}